\documentclass{aa}  

\usepackage{graphicx}
\usepackage{amsmath,amssymb}
\usepackage{txfonts}
\usepackage{hyperref}
\usepackage{xcolor}
\usepackage{kvoptions}

\usepackage{tikz}
\usepackage{verbatim}
\usetikzlibrary{arrows,shapes}

\begin{document} 

\newcommand{\vinfvesc}{$\varv_\infty/\varv_\mathrm{esc,\Gamma_e}$}

   \title{The winds of OBA hypergiants and luminous blue variables}
   \subtitle{Dynamically consistent atmosphere models reveal multiple wind regimes}
   
   \author{
        M.~Bernini-Peron\inst{\ref{inst:ari}}
        \and
        A.A.C.~Sander\inst{\ref{inst:ari},\ref{inst:iwr}}
        \and 
        G.N.~Sabhahit\inst{\ref{inst:aop}}
        \and 
        F.~Najarro\inst{\ref{inst:cab}}
        \and
        V.~Ramachandran\inst{\ref{inst:ari}}
        \and
        J.S.~Vink\inst{\ref{inst:aop}}
    }

   \institute{
           {Zentrum f{\"u}r Astronomie der Universit{\"a}t Heidelberg, Astronomisches Rechen-Institut, M{\"o}nchhofstr. 12-14, 69120 Heidelberg\label{inst:ari}\\
        \email{matheus.bernini@uni-heidelberg.de}}
    \and 
          {Interdisziplin{\"a}res Zentrum f{\"u}r Wissenschaftliches Rechnen, Universit{\"a}t Heidelberg, Im Neuenheimer Feld 225, 69120 Heidelberg, Germany\label{inst:iwr}}
     \and
           {Armagh Observatory and Planetarium, College Hill, Armagh BT61 9DG, N. Ireland\label{inst:aop}}
    \and
           {Departamento de Astrof\'{\i}sica, Centro de Astrobiolog\'{\i}a, (CSIC-INTA), Ctra. Torrej\'on a Ajalvir, km 4,  28850 Torrej\'on de Ardoz, Madrid, Spain\label{inst:cab}}
   }

   \date{Received December 23, 2025; accepted February 21, 2026}
 
  \abstract
    {
    OBA hypergiants are evolved massive stars with notable wind features in their optical spectrum. Located at the cool end of the line-driven wind regime, many of these objects are candidate luminous blue variables (LBVs) and presumably close to the Eddington limit.
    Despite representing a rather short-lived stage in the life of massive stars, they strongly affect their surroundings and subsequent stellar evolution via their high wind mass loss.
    }
    {
    We investigate the atmospheric conditions and mechanisms that produce the winds and spectral appearance of hypergiants in the OBA temperature regime, thereby also covering typical temperatures of minimum and maximum phases of noneruptive LBVs.
    } 
    {
    We used the hydrodynamically consistent version of the atmosphere code \texttt{PoWR}  to produce a sequence of atmosphere models with a classical Eddington parameter $\Gamma_\mathrm{e} \sim 0.4$ and moderate turbulent pressure, in line with typical parameters obtained in the regime of cool supergiants and hypergiants. We varied the effective temperature at the inner boundary from $\sim$12.5~kK to $\sim$38.0\,kK. Our main model series was calculated at solar metallicity, with a few additional tests performed at lower metallicity.
    }
    {
    We found a complex pattern of the mass-loss rate as a function of temperature in the hypergiant regime, with regions of higher and lower rates associated with different types of wind solutions, which we call ``dense'' and ``rarefied''. We found known hypergiants and LBVs with spectra resembling the synthetic spectra of our models for all of the wind-solution regions.
    Around the temperatures where \ion{Fe}{IV} recombines to \ion{Fe}{III} and \ion{Fe}{III} recombines to \ion{Fe}{II}, we found a bistability of solutions and sharp increases in the mass-loss rate. In addition, we found decreases when the leading Fe ion begins to change at the wind onset, indicating that the wind solutions switch from dense to rarefied when the driving opacity is insufficient.
    The resulting velocity fields in the different regions reflect the different atmosphere solutions, where the rarefied solutions agree with the empirical relation of the terminal velocity and effective temperature.
    }
    {
    Our results demonstrate the existence of the first and second bistability jump and their association with Fe recombination in the hypergiant regime. However, the jumps are embedded in an overall more complex behavior that agrees but poorly with any existing mass-loss recipe. Toward cooler temperatures, the role of turbulent pressure in the wind acceleration becomes increasingly important, in particular in the rarefied regimes.
    Combining our results with other recent modeling efforts, we suggest that the switch between rarefied and dense regimes only occurs within a certain proximity to the Eddington limit, while otherwise, only either dense or rarefied solutions will be obtained. Calculations exploring a wider parameter regime and incorporating a more sophisticated treatment of radiatively driven turbulence are necessary to further test this hypothesis.
    }
    {}


   \maketitle

\section{Introduction}
\label{sec:intro}

Mass loss is a key ingredient that can shape the entire evolution of a massive star ($M \gtrsim 8$\,$\mathrm{M_\odot}$) since its birth as an O or early-B star until their eventual core collapse \citep[see, e.g.,][for recent reviews]{Langer2012,Smith2014,Vink2022}.
While the post-main-sequence stage represents only $\sim$10\% of a star's lifetime, dramatic changes in the stellar properties, such as the surface chemical composition, luminosity, and temperature, can be observed. These changes considerably affect the resulting stellar wind properties, such as the mass-loss rate $\dot{M}$, which in turn is a major ingredient for the subsequent evolution of a star. Therefore, a quantitative understanding of the interplay between the stellar and wind properties is key to better connecting the observed stages of massive post-main-sequence stars and predict their evolutionary fates.

Toward the cool end of the line-driven wind regime reside late-B and A supergiants and hypergiants\footnote{We use the definition for hypergiants by \cite{Clark+2012} based exclusively on spectral morphology \citep[cf.][]{Negueruela+2024}.}. Many of these are luminous blue variables (LBVs) or LBV candidates. In its raw definition, the LBV phenomenon as such is not an evolutionary stage, but an observational umbrella term \citep{Conti1984} that encompasses several different objects and phenomena, namely eruptive variables (e.g., $\eta$ Car), P-Cygni supergiants (e.g., P Cyg and HD~316285), and S-Dor variables (e.g., S Dor and MWC\,930). 

Members of this class of objects, which frequently produce nebulae around themselves, are expected to play a pivotal role in the evolution of massive stars as they are likely the gateway to stages such as Wolf-Rayet stars (WRs) or core-collapse supernovae (CCSNe), especially of type IIn \citep[e.g.][]{Kotak-Vink2006,Trundle+2008,2020+Nyholm}. 
Moreover, they are expected to be the second main source of metal enrichment of their host galaxies \citep[e.g.,][]{Agliozzo+2021}.
They are also producers of complex carbon molecules because their temperatures and mass-loss rates are sufficiently high and they are highly variable. This is extremely relevant for astrobiology \citep{Arun+2025}.

This zoo of stars spans a wide range of effective temperatures and luminosities \citep[$T_\mathrm{eff} = 6\dots40$\,kK; $\log L = 5.0 \dots 6.5$; e.g.,][]{Smith+2019}. This parameter range encompasses regions in which theoretical models traditionally predict a discontinuous behavior in the wind parameters, which is attributed to the also found discontinuities in a bistability of solutions found originally by \citet{Pauldrach_Puls1990} when modeling P\,Cyg under the framework of the ``modified CAK theory'' \citep[][hereafter mCAK]{Castor+1975,Pauldrach+1986, Friend-Abbott1986}.
\citet{Lamers+1995} later found discontinuities in the ratio of the terminal velocity $\varv_\infty$ and the effective escape velocity,
\begin{equation}
  \label{eq:vescg}
  \varv_\mathrm{esc, \Gamma_\mathrm{e}} = \sqrt{\frac{2 G M_\star}{r} \left(1 - \Gamma_\text{e}\right)},
\end{equation} 
for a sample of OBA supergiants.

Incorporating this discontinuity in \vinfvesc, modeling efforts in a broader parameter regime using a Monte Carlo approach \citep{Vink+1999} yielded two particular sharp increases (jumps) in the behavior of $\dot{M}$ as a function of decreasing effective temperature ($T_\text{eff}$): A first jump at   $T_\mathrm{eff} \sim 25$\,kK, the so-called first bistability jump (1-BiSJ), which can be traced back to a recombination from \ion{Fe}{IV} to \ion{Fe}{III} in the atmosphere, and a second jump at $T_\mathrm{eff} \sim 10$\,kK, the second bistability jump (2-BiSJ), which is attributed to a recombination from \ion{Fe}{III} to \ion{Fe}{II}. These are also included in the mass-loss description from \cite{Vink+2001}, which is widely employed in stellar evolution. 
Further, the jumps have been found independently of any assumption on \vinfvesc\ \citep[e.g.,][]{Mueller-Vink2008, Vink-Sander2021}.
Commonly, only the 1-BiSJ is explicitly incorporated, while the switch to the $\dot{M}$ treatment for cooler stars approximately incorporates the effect of the 2-BiSJ when a recipe with high mass loss is used, such as that by \citet{deJager+1988}.

Recent multiwavelength quantitative spectroscopic studies of BSGs challenged the 1-BiSJ scenario \citep[e.g.,][]{Bernini-Peron+2023, Bernini-Peron+2024, Verhamme+2024, Alkousa+2025, deBurgos+2024}. They did not find signs of discontinuous increases and reported a rather constant trend at best that was better captured by the $\dot{M}$ prescriptions of \cite{Krticka+2024}. Moreover, the discontinuity in \vinfvesc\ is also contested \citep{Crowther+2006,Markova_Puls2008, Alkousa+2025}. 
However, there is some evidence for the 1-BiSJ in LBVs based on analyses of AG Car throughout its variability \citep[e.g.,][]{Vink-deKoter2002, Groh+2011, Groh+2011b} and dynamical structure modeling \citep{Grassitelli+2021}.

The scientific panorama for the 2-BiSJ remains very unexplored spectroscopically because the UV spectral coverage of late-B and A supergiants is far sparser, and modeling this parameter regime is challenging. Therefore, hardly any investigations were dedicated specifically to the wind behavior in this regime{\color{black}. In yellow supergiants and hypergiants (YSGs and YHGs), the 2-BiSJ is thought to trigger dust shells around some of these stars \citep[e.g.,][]{Koumpia+2020}}. Using a grid of \texttt{CMFGEN} models with global wind consistency, \cite{Petrov+2016} found that the 2-BiSJ is associated with the change in ionization of \ion{Fe}{III} to \ion{Fe}{II} for models with a high luminosity-to-mass ratio ($L/M$). \citet{Sabhahit+2026}, using a grid of fully hydrodynamically consistent models, also found an increase in $\dot{M}$ that was driven by \ion{Fe}{II} for models with high $L/M$. However, their investigation was motivated by the enhanced $\dot{M}$ of very massive stars, and thus, their grid in this region is sparse.

We use hydrodynamically consistent atmosphere models to show that the behaviour of mass loss and other wind properties of stars closer to the Eddington limit in the Hertzsprung gap region (usually occupied by hypergiants and LBVs) is indeed complex. This is strongly tied to the interplay of changes in the ionization of \ion{Fe}{}, H, and He, and to the action of the star's atmospheric turbulent pressure.

In Sect.\,\ref{sec:methods} we briefly describe our tools, approach, and assumptions. In Sect.\,\ref{sec:results} we briefly illustrate the resulting optical spectra and their resemblance to those of known LBVs and BHGs before we present the trends in mass-loss rate and wind structure we obtain.
We then discuss the implications of our results in Sect.\,\ref{sec:discussion} before we draw our conclusions in Sect.\,\ref{sec:conclu}.

\section{Methods}
\label{sec:methods}

\begin{figure}
\centering
\includegraphics[width=1.0\linewidth]{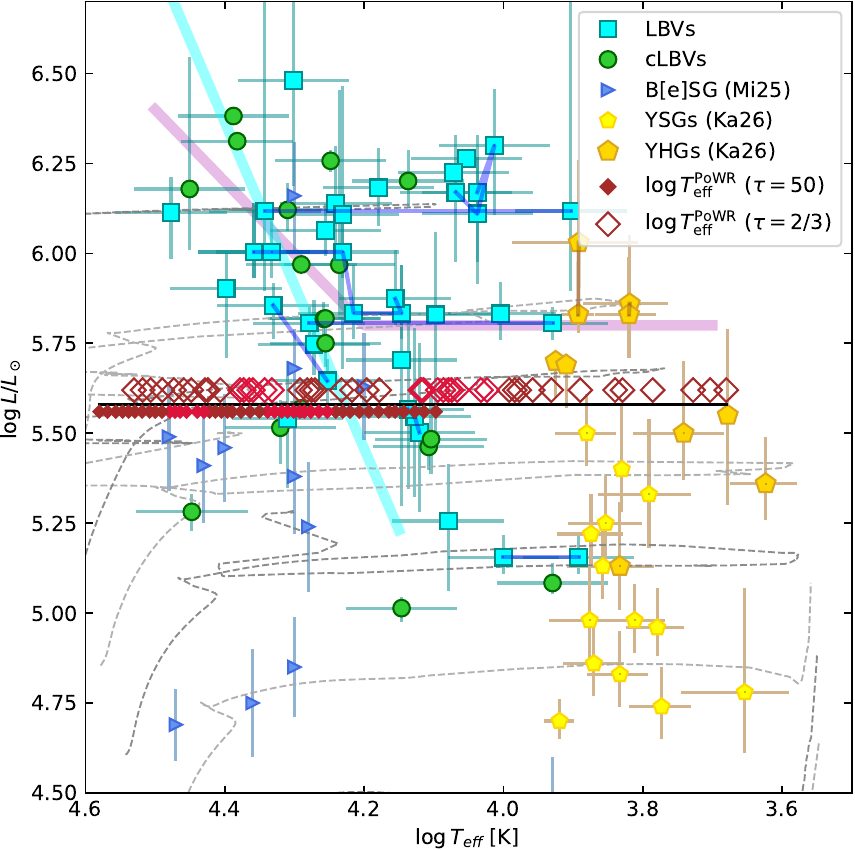}
  \caption{
    Hertzsprung-Russell diagram depicting our model sequence against LBVs (cyan squares), B[e]SG (blue triangles), OBA hypergiants or LBV candidates (green circles), YSGs (small yellow pentagons), and YHGs (large gold pentagons) from the literature (see Appendix Sect.\,\ref{sec:distance-lum}). The blue lines connecting the LBV data indicate the same star in different epochs. Likewise, the dark orange lines connect YHGs in different epochs. The luminosities are updated to distances mostly based on Gaia DR3 \citep{BailerJones+2021, GaiaCollab+2023}. The black line indicates the luminosity of our sequence. The small diamonds below indicate $T_*$, and the large open diamonds indicate $T_\mathrm{2/3}$. The thin dotted lines indicate the Galactic evolutionary tracks by \cite{Ekstroem+2012}. The thick cyan line indicates the LBV instability strip \citep{Groh+2011}, and the thick violet line indicates the HD limit \citep{Humphreys-Davidson1979}.
  }
    \label{fig:HRD}
\end{figure}

We used the stellar atmosphere code \texttt{PoWR$^\textsc{hd}$} \citep{Sander+2018}  to produce a fine-spaced sequence of hydrodynamically consistent models of massive stars with moderate mass and luminosity that were still close to the Eddington limit.
We selected the option to fix all stellar parameters to \texttt{PoWR$^\textsc{hd}$} when we computed the velocity field (including\ $\varv_\infty$) and $\dot{M}$.

We fixed the mass (20.6\,$M_\odot$), luminosity ($\log L/\mathrm{L_\odot} = 5.58$; representative values of that lead to an $L/M$ typical for hypergiants and LBVs), and a mildly enriched He abundance (with a He mass fraction $X_\mathrm{He} = 0.33$) as well as the same CNO composition as in \cite{Bernini-Peron+2025}. 
Motivated by time-dependent hydrodynamical simulations of \cite{Driessen+2019}, who found that stars below $T_\mathrm{eff} < 20$\,kK have lower clumping, we fixed the volume filling factor $f_\mathrm{vol} = 0.33$ at all depth points.

We adopted a depth-independent turbulence velocity ($\varv_\mathrm{turb} = 20$\,km\,s$^{-1}$), as is typically employed in atmospheres of LBVs and even some BSGs and BHGs \citep[e.g.,][]{Groh+2009}. However, we note that this value might be an underestimation, especially for the hotter models of our sequence, as recent multidimensional simulations of atmospheres for stars with high $\Gamma_\mathrm{e}(R_\mathrm{crit})$ \citep{Moens+2025} predict much higher values of $\varv_\mathrm{turb} \sim 150\,\mathrm{km\,s^{-1}}$. As we aim for a homogeneous sequence of models isolating the effect of a $T_\star$ variation, we kept $\varv_\infty$ (and other parameters) fixed.

In the atomic physics, we considered the elements H, He, C, N, O, Si, S, Al, P, and Mg, and the iron-group elements (G), which, albeit largely dominated by Fe, conflate levels and transitions from Sc to Ni. In Appendix\,\ref{sec:atomic}, we disclose the considered levels and transitions for each element.

\begin{figure}
\centering
\includegraphics[width=1.0\linewidth]{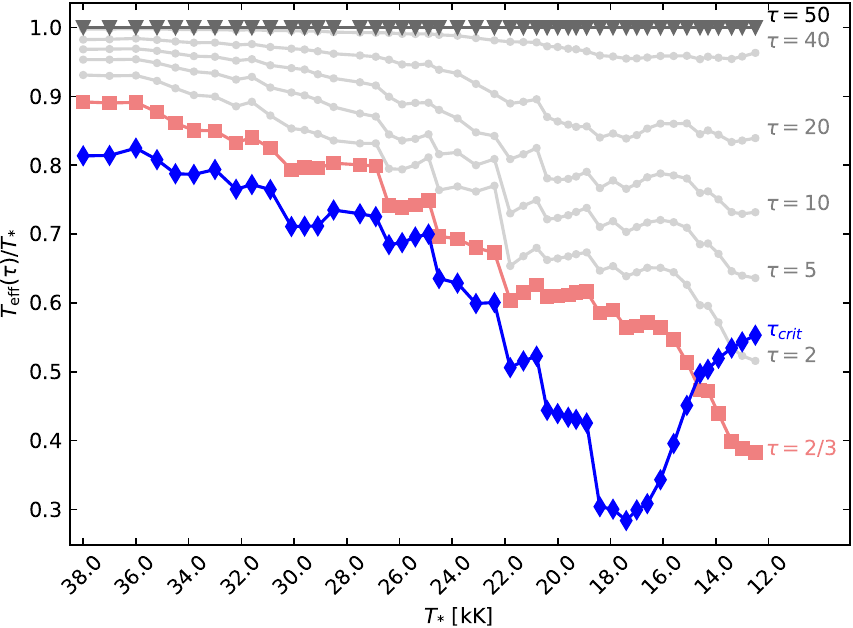}
  \caption{
  Effective temperature at different optical depths $\tau$, normalized by $T_\star = T_\mathrm{eff}(\tau=50)$ for each point of the sequence. The pink curve indicates $T_{2/3}$, and the blue curve indicates $T_\mathrm{eff}$ at the critical point.
  }
    \label{fig:teff-opt-depths}
\end{figure}

In our main model sequence, we only varied the temperature at the inner boundary $T_\star = T_\mathrm{eff}(\tau = 50)$, where $\tau$ is the Rosseland continuum optical depth, from 13~kK to 38~kK. The photospheric temperature, that is, $T_\mathrm{eff} (\tau = 2/3)$ (hereafter simply $T_{2/3}$) is an output quantity that depends on the resulting $\dot{M}$.
As our model sequence has high $\Gamma_\mathrm{e}$ values, the models' atmospheres are expanded, such that the difference between $T_\star$ and $T_{2/3}$ amounts to $\sim$7~kK. In Fig.\,\ref{fig:HRD} we show the location of our model sequence, indicating $T_\star$ and $T_{2/3}$, in the Hertzsprung–Russell diagram (HRD) together with a set of known objects in the hypergiant and LBV domain. All comparison data were taken from the literature, although the luminosities were updated when newer distance information from Gaia DR3 was available (see Appendix~\ref{sec:distance-lum}).

A more direct illustration of the difference between $T_\star$ and $T_{2/3}$ at different optical depth is shown in Fig.\,\ref{fig:teff-opt-depths}. Despite some scatter on small scales, the atmospheres clearly become more extended with cooler $T_\star$, and the difference between $T_\star$ and $T_\mathrm{eff} (\tau < 50)$ increases. For the coolest models, the effective temperature of the critical point becomes lower than $T_{2/3} = T_\mathrm{eff} (\tau = 2/3)$, meaning that the wind stars to launch beneath the layers forming the continuum. This is a signature of an optically thick wind and is known in particular from models describing classical WRs \citep[e.g.,][]{Sander+2020,Sander+2023,Lefever+2023, Lefever+2026}.

\section{Results}
\label{sec:results}

\subsection{Spectral appearance}

\begin{figure*}
\centering
\includegraphics[width=1.0\linewidth]{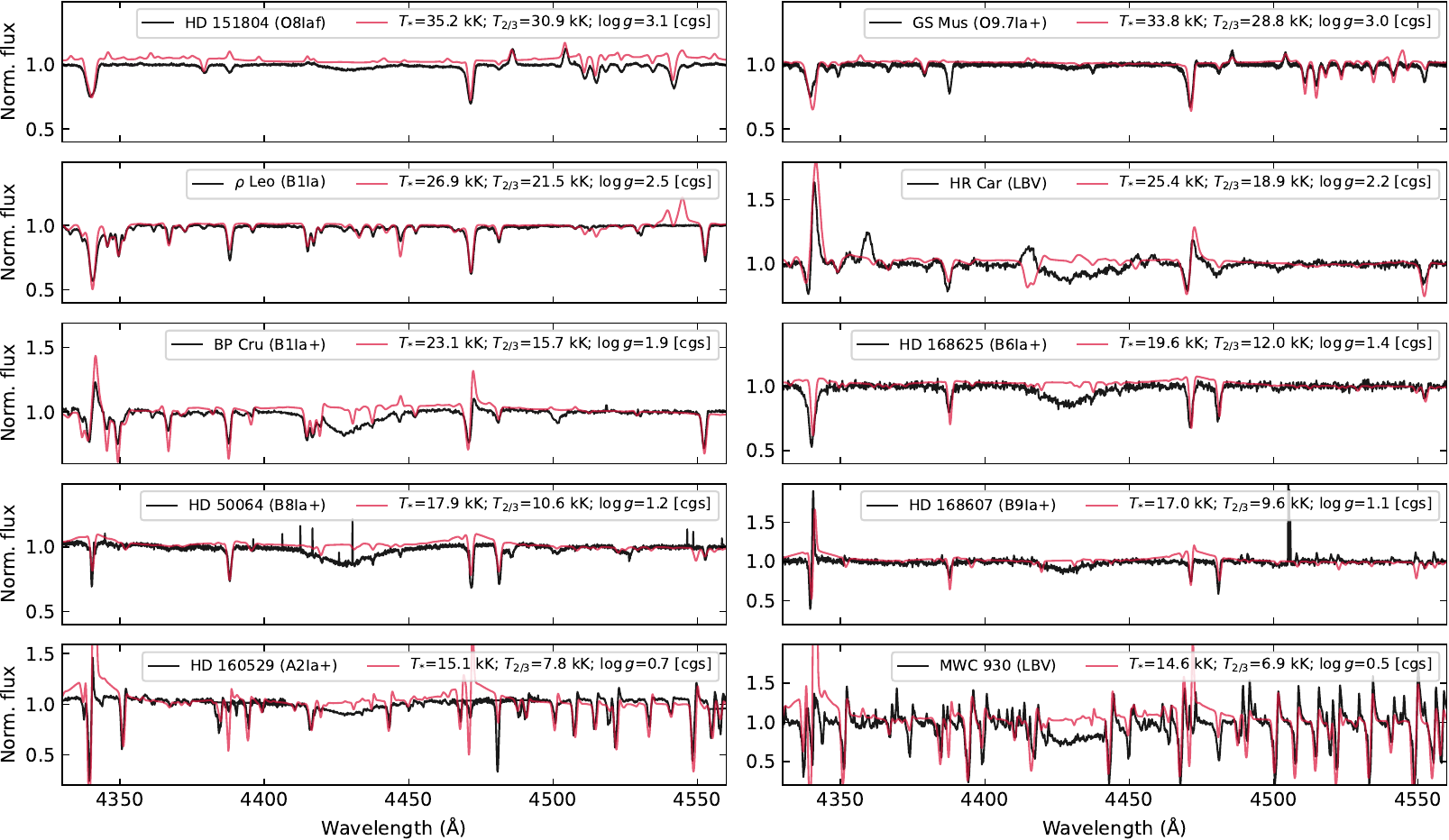}
  \caption{
    Nontailored comparison between the output spectra of some of the models (red lines) and observed spectra (black lines) of OB supergiant/hypergiants and LBVs.
  }
    \label{fig:spectrum-similarity}
\end{figure*}

Our \texttt{PoWR$^\textsc{hd}$} models are not specifically designed to fit any particular star and have a fixed luminosity. However, as illustrated in Fig.\,\ref{fig:spectrum-similarity}, we found that the normalized synthetic spectra of our models generally resemble spectra of existing hypergiants and LBVs in their temperature domain. This indicates that many hypergiants, LBVs, and even some supergiants that spectroscopically appear to be very different from each other might belong to this $L/M$ regime.
Moreover, this is an important sanity check for the robustness of the atmosphere structures and indicates that our sequence reflects their physics in the hypergiant regime.

\subsection{The behavior of the mass-loss rate}

\begin{figure}
    \centering
    \includegraphics[width=1\linewidth]{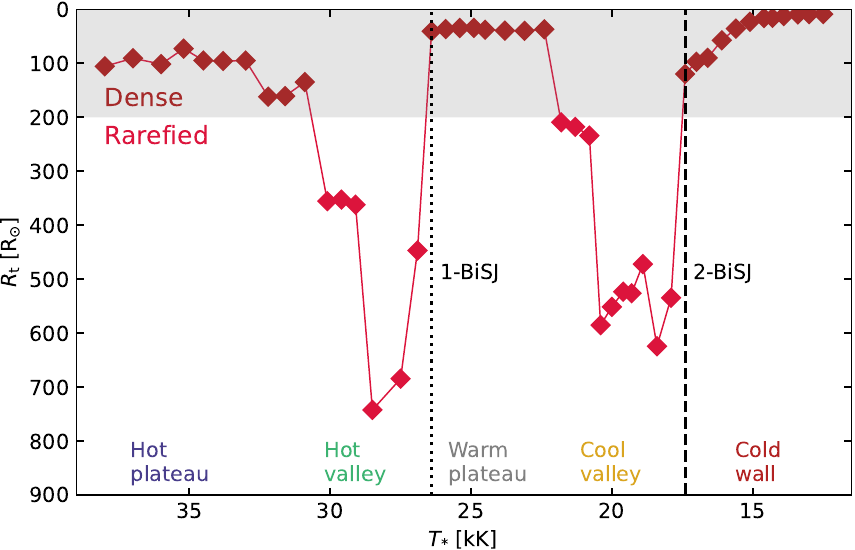}
    \caption{Transformed radius ($R_\mathrm{t}$) vs. $T_\ast$. Models with $R_\mathrm{t} < 200$\,R$_\odot$ are in the dense-wind regime, i.e., with higher $\dot{M}$ and optical spectra crowded with emission and P~Cygni features (shaded region). The colored texts in the lower part of the diagram name the wind-solution regions identified in the $T_\star$ sequence. The 1-BiSJ and the 2-BiSJ are indicated by the dotted and dashed lines in the transitions from a valley to a plateau (i.e., increase in $\dot{M}$) toward lower $T_\star$.}
    \label{fig:airy-dense-Rt}
\end{figure}

\begin{figure*}
\centering
\sidecaption
\includegraphics[width=12cm]{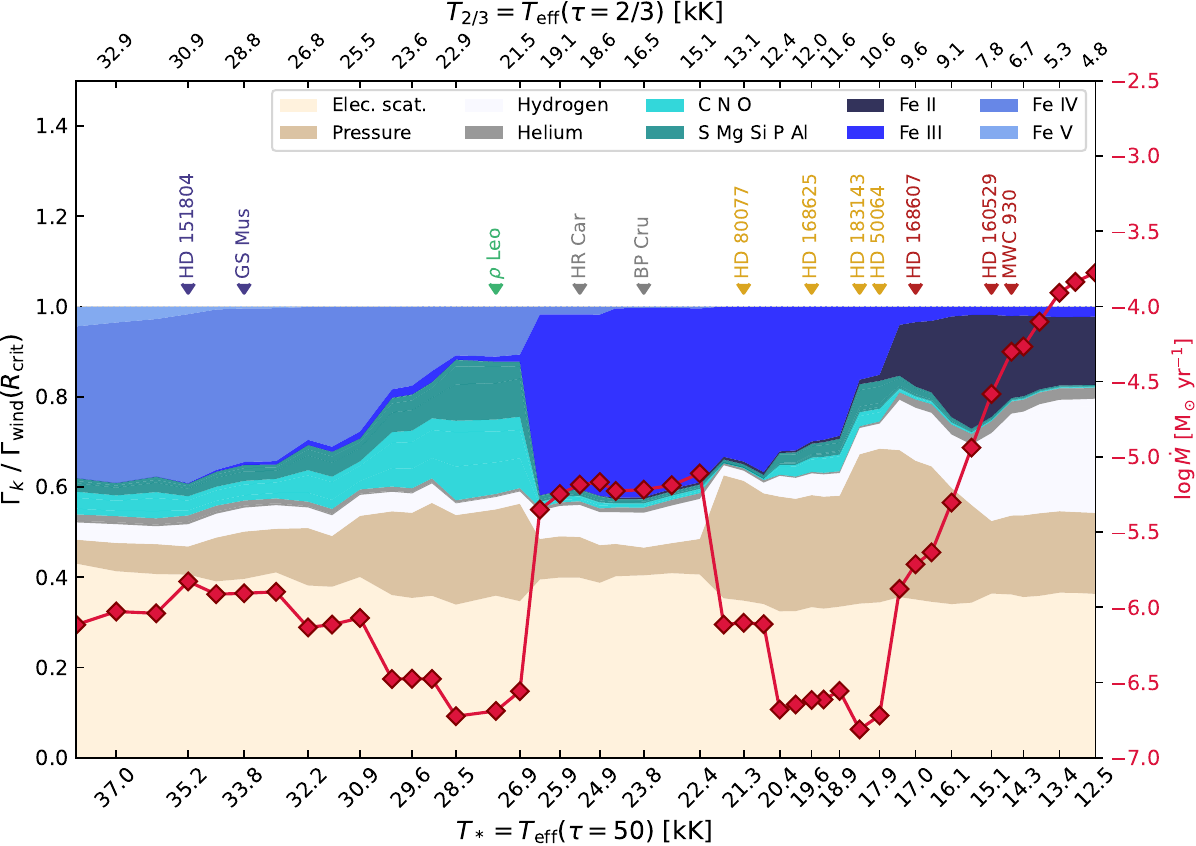}
  \caption{
  Contributions of different processes, elements, and ions to the wind launching at the critical point (stacked area plot) and corresponding mass-loss rates (red line) for different inner boundary temperatures ($T_*$). From bottom to top, the light beige area indicates the electron-scattering acceleration, dark beige shows the pressure, and smoke white indicates \ion{H}{I} continuum driving. The teal tones indicate the contribution of CNO and other metals, and the shades of blue indicate the contributions of different Fe ions.
}
    \label{fig:ion_acc_mdot}
\end{figure*}

Figs.~\ref{fig:airy-dense-Rt} and \ref{fig:ion_acc_mdot} show a complex pattern for the derived mass-loss rate as a function of $T_\star$. The transformed radius,
\begin{equation}
  \label{eq:rt}
   R_\mathrm{t} = R_\star \left(\frac{\varv_\infty}{\dot{M}\sqrt{f_\mathrm{cl}}} \cdot \frac{10^{-4 \,} \mathrm{M_\odot \,yr^{-1}}}{2500\, \mathrm{km\,s^{-1}}} \right)^{2/3},
\end{equation} 
describes the ratio of the emission measure of the wind and the surface area at $R_\ast$, and in the case of our $T_\mathrm{*}$ sequence, it separates the geometrical effect from the wind behavior. Two main types of wind solutions can be identified: dense-wind solutions with $R_\mathrm{t} > 200$~$\mathrm{R_\odot}$, and rarefied-wind solutions\footnote{We did not label them ``thick'' and ``thin'' as they do not coincide with the wind-optical depth $\tau_\text{F}$ crossing unity. The thick solutions have a higher $\tau_\text{F}$, but there is no uniform transition value among the sequence.} with smaller $R_\text{t}$ (see Fig.\,\ref{fig:airy-dense-Rt}). Significant jumps to higher $\dot{M}$ appear around the temperatures associated with the traditional bistability jumps, but these jumps are not the only abrupt transitions. Instead, Fig.~\ref{fig:ion_acc_mdot} shows a more complex pattern with jumps to higher and also drops to lower $\dot{M}$. In the following, we refer to these as valleys in $\dot{M}(T_\star)$.

In the background of Fig.\,\ref{fig:ion_acc_mdot}, we show a stacked area plot indicating the contributions to the total outward pushing forces
\begin{equation}
  \Gamma_\text{wind} := \Gamma_\text{rad} + \Gamma_\text{press} 
\end{equation}
at the critical point $R_\text{crit}$, with $\Gamma_\text{rad}$ summarizing the radiative line and continuum accelerations, while $\Gamma_\text{press}$ includes the acceleration due to gas and turbulent pressure gradients, all normalized to gravity $g$.
The temperatures associated with the two classical BiSJs are clearly associated with the change in the ionization of \ion{Fe}{}. 
However, the other end of the valleys, that is, the decline of $\dot{M}$ to the low values before the jumps, is less abrupt and seems to be associated with the gradual increase in the contribution of other elements and the pressure term. In turn, this seems to be indirectly caused by a slow transition of the dominance of different \ion{Fe}{} ions. A similar behavior was recently mapped for stars close to the Eddington limit \citep[][]{Lefever+2025,Sabhahit+2026}, but for a much higher luminosity regime. 

\begin{figure}
    \centering
    \includegraphics[width=1\linewidth]{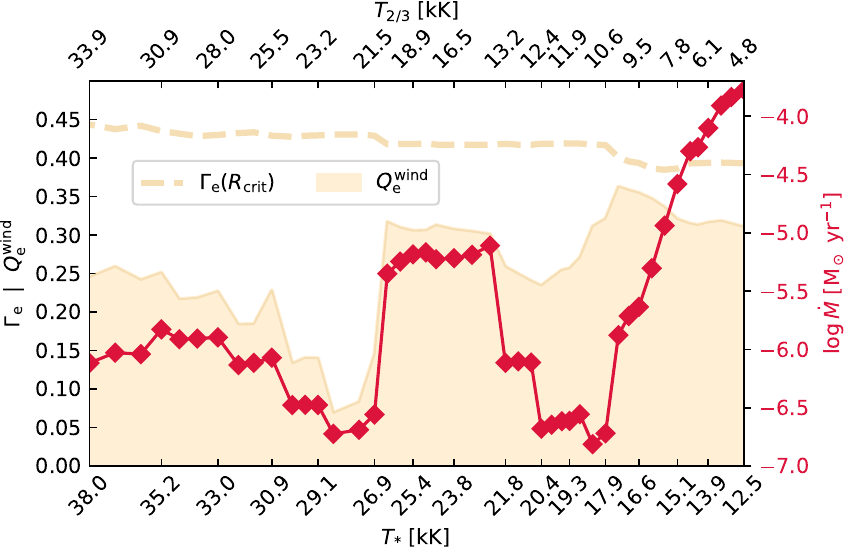}
    \caption{Comparison between the trend of mass-loss rates (red curve) and the classical Eddington parameter at the critical point $\Gamma_\mathrm{e}(R_\mathrm{crit}$) (thick dotted curve) and the total contribution of the electron scattering to the work ratio $Q_\mathrm{e}^\mathrm{wind}$ (shaded area). }
    \label{fig:gammae_mdot}
\end{figure}

A further inspection of the relative contributions to $\Gamma_\text{wind}$ in Fig.\,\ref{fig:ion_acc_mdot} shows that radiatively driven processes dominate the wind acceleration. However, the sheer proximity to the Eddington limit of the model sequence ($\Gamma_\text{e} \sim 0.4$) leads to the effect that in all models, less than half of the acceleration at the critical point is provided by line opacities. In addition to a large contribution from free-electron (i.e., Thomson) scattering, the continuum opacity of hydrogen, and the acceleration due to gas and turbulent pressure gradients, need to be taken into account. In the following, we discuss the different features in the mass-loss trend. We also consider for this the work ratios
\begin{equation}
    Q_k = \int_{R_\star}^{\infty} a_{\mathrm{rad},k}\,\mathrm{d}r\, \Big/ \int_{R_\star}^{\infty} \left(\varv\frac{\mathrm{d}\varv}{\mathrm{d}r} + \frac{GM_\star}{r^2}\right) \mathrm{d}r\text{,}
\end{equation}
where the $a_{\mathrm{rad},k}$ is the radiative acceleration due to the respective contribution $k$ (e.g., electron scattering, gas pressure, Fe, H, and CNO). In Fig.\,\ref{fig:gammae_mdot} we plot $\Gamma_\text{e}(r_\mathrm{crit})$ and $Q_\text{e}$, which shows that the valleys in $\dot{M}$ match regions of low  $Q_\text{e}$, that is, a low global contribution of the electron scattering and thus a high amount of line driving.

\paragraph{The hot dense-wind plateau}
The hottest models in the sequence have moderately high $\log \dot{M} \sim -6$ and probe the temperature range of OHGs. The winds in this region are largely driven by $\Gamma_\mathrm{e}$- and \ion{Fe}{IV}. Pressure and other elements contribute up to $\sim$25\% to launch the wind.

The obtained terminal velocities match the $\varv_\infty$ determined by \cite{Prinja+1990} for GS\,Mus \citep[O9.7Ia+,][]{Carretero-Castrillo+2023}, one of the few Galactic OHGs that has a UV spectrum.
The $T_{2/3}$ and $\log g(\tau=2/3)$ also match the results obtained by \cite{Holgado+2020}, which results in a very similar spectral appearance.
We also find a similar spectral appearance of HD\,151804 (formally an O8Iaf) in the optical. In the UV, we find an apparently slow wind. It is possible that elements like Ar, which are shown to affect $\varv_\infty$ in earlier O stars and are not included in our sequence, might start to play a role.
While we  did not perform a dedicated spectral analysis, we scaled the flux of our models to the observed SED (see appendix Sect.\,\ref{sec:OHG-lum}) to determine the luminosities of GS\,Mus and HD\,151804 ($\log L/\mathrm{L_\odot} \sim 5.2$ and $\log L/\mathrm{L_\odot} \sim 6.0$) and placed them in the HRD (see Fig.\,\ref{fig:HRD}).

\paragraph{The hot rarefied-wind valley}
As \ion{Fe}{IV} slowly loses dominance, the available opacity is insufficient to maintain a deeper wind-launching associated with a higher mass-loss rate. With decreasing $\tau_\text{crit}$, CNO and other metal lines contribute quite significantly to the wind launching, as does the pressure term. In this regime, the winds reach $\sim$1000~km~s$^{-1}$, which resembles winds of BSGs ($\sim$B1~Ia) on the verge of the predicted classical 1-BiSJ \citep{Vink+2001}. Spectroscopically, the models in this range and the B1~Ia supergiant $\rho$~Leo match as well.

\paragraph{The classical first jump and the warm dense-wind plateau}
At $T_* \sim 27$~kK ($T_{2/3} \sim 22$~kK), the sharp increase in $\dot{M}$ is directly associated with the increase in \ion{Fe}{III} that contributes to the wind driving. This is the 1-BiSJ already seen in the Monte Carlo models by \citet{Vink+1999, Vink-Sander2021}, albeit with shifts in the temperature. In appendix \ref{app:bistable} we show that we found a bistability of solutions in this regime transition, but we only selected one of them for clarity of the overall plots, such as Fig.\,\ref{fig:ion_acc_mdot}. Toward cooler temperatures, the regime of high $\dot{M}$ persists until $T_\star \sim 23$~kK ($T_{2/3} \sim 15$~kK). In contrast to the results by \cite{Vink+1999}, we found no gradual decrease in $\dot{M}$ after the jump, but a plateau. As for the hot plateau before, the contributions other than from Fe and $\Gamma_\mathrm{e}$ only amount to $\sim$20\%.

The wind solution we obtained for this region is very similar to the solution obtained for the \texttt{PoWR$^\textsc{hd}$} model of $\zeta^1$~Sco \citep[B1.5Ia+][]{Bernini-Peron+2025}. The corresponding spectra for models in this regime resemble the spectrum of BP~Cru (B1Ia+/HMXB), a hypergiant with similar properties to $\zeta^1$~Sco \citep{Clark+2012}. 
We also note that it resembles the spectrum of HR Car in 2017, suggesting that the star is currently in this dense-wind regime. In this temperature range, we also find LBVs such as P~Cyg (B1Ia+/LBV), whose spectra are much more extreme, which suggests either a considerably different clumping structure or a higher $L/M$ ratio. In contrast, we also see stars like HD190603 (B1.5Ia+) and  $\zeta^1$~Sco, which have less prominent P Cygni features (see Sect.\,6.3 in \cite{Bernini-Peron+2025} for a discussion on the effects of clumping in the spectra of early-type BHGs).

\paragraph{The cool rarefied-wind valley}

At $T_\star \sim22$\,kK ($T_{2/3} \sim 15$\,kK), the mass-loss rates drop by more than an order of magnitude.  
This behavior is the opposite of what was predicted for the classical BiSJ and is not associated with any major changes in the ions (in particular Fe) driving the wind. Nonetheless, the wind solution changes and $\tau_\text{crit}$ decreases again because the \ion{Fe}{III} opacities are no longer sufficient to launch the wind at similar radii as obtained for the solutions of the warm dense-wind plateau. \ion{Fe}{II} is not sufficiently populated to take over the wind launching, however. Instead, the pressure term supports (quasi-)hydrostatic equilibrium, until an outflow is launched farther out, at the expense of a considerably reduced $\dot{M}$. Unlike the hot rarefied-wind valley, non-Fe metal lines do not significantly contribute to the wind launching ($\lesssim10$\%), while the contribution of the pressure term is even stronger.

The synthetic model spectra in this regime resemble the appearance of several late-type BHGs (B2$\dots$B9), for instance, HD80077, HD183143, and HD168625.
Notably, many of the known BHGs in this temperature regime ($T_{2/3} \sim 14\dots10$~kK) also display the obtained 
weak hypergiant characteristics, that is, H$\beta$, H$\gamma$, H$\delta$, with possibly He and metal lines in P Cygni or emission. 
The hypergiant status is questioned for some of those stars because H$\beta$ is variable and often fully in absorption \citep[see][]{Negueruela+2024}.
Our finding of the cool rarefied-wind valley indicates that these weaker characteristics might be a natural outcome of the opacity situation in the atmosphere and do not reflect a lower $L/M$. This would mean that earlier-type BHGs such as HR\,Car, BP\,Cru, or $\zeta^1$~Sco are not necessarily closer to the classical Eddington limit, but rather have more favorable wind conditions.

\paragraph{The classical second jump and the cold dense-wind wall}

Toward even lower temperatures, $\dot{M}$ increases significantly at $T_\star \sim 17$\,kK ($T_\mathrm{2/3} = 10$\,kK), which is associated with the change from \ion{Fe}{III} to \ion{Fe}{II} in the wind driving. The higher wind density causes the model spectra in this regime to resemble AHGs and LBVs that have P~Cygni and emission features, such as HD160529 and the LBV MWC~930, crowded with \ion{Fe}{II} profiles in the case of the coolest models.
Approximately similar in $T_\mathrm{2/3}$, this essentially represents the 2-BiSJ predicted by \cite{Vink+1999} and \citet{Petrov+2016}. The jump also occurs in the LBV models by \citet{Vink-deKoter2002}, albeit at temperatures more similar to our $T_\star$ value. The 1-BiSJ also has a bistability of solutions at the onset of the cold dense-wind wall. We illustrate this further in Appendix \ref{app:bistable}.

Notably, the 2-BiSJ in \citet{Petrov+2016} only appeared in models closer to the classical Eddington limit ($\Gamma_\mathrm{e} = 0.39$, $\log L/\mathrm{L_\odot} = 5.5$, $M = 20$\,M$_\odot$, and $Y = 0.25$) when they transitioned from $T_{2/3} = 10.0$\,kK to $T_{2/3} = 8.8$\,kK. The jump in their analysis appears in a very similar parameter space and between very similar $\dot{M}$ values ($-7.04$ to $-5.70$) as those that we obtain ($-6.7$ to $-5.8$). 
Interestingly, this similarity in outcome does not coincide with the same driving explanation. \citet{Petrov+2016} reported that \ion{Fe}{II} causes $\sim$70\% of the total work ratio, while in our case, \ion{Fe}{II} contributes only up to $\sim10$\% (first panel of Fig.\,\ref{fig:grand-acc-mdot}) and most is contributed by electron scattering and pressure. 
Moreover, when we consider the total work ratio $Q_\mathrm{wind}$ for our models in the cold dense-wind wall, all the Fe ions contribute similarly. In addition, the total Fe contribution to $Q_\mathrm{wind}$ is equivalent to that of H and much smaller than that of electron scattering or pressure.

In the models from \citet{Petrov+2016}, the 2-BiSJ marks the last point of their model sequence. In their models, \ion{Fe}{II} is not fully recombined at $T_{2/3} = 8.8$\,kK, raising the question of the further trend of $\dot{M}$ versus\ temperature. 
Our model sequence extends slightly farther, and $\dot{M}$ continues to increase together with the contribution of \ion{Fe}{II}.
During the S~Doradus cycle, these high values of $\dot{M}$ might explain nebulae around many LBVs and hypergiants, including those that currently do not present themselves as bona fide LBVs. Ahigh $\dot{M}$ past the 2-BiSJ like this might also explain the signatures of type IIn SNe \citep[e.g.,][]{Kotak-Vink2006, Trundle+2008}.
Within the capability of our current simulations, we did not reach a stage in which the trend in $\dot{M}$ turns over into another plateau or valley. However, we found an indication of a possible stagnation as the increase in $\dot{M}$ becomes smaller for the coolest models in the series, which was also suggested by the $\dot{M}$ behavior found by \cite{Vink-deKoter2002}.

\subsection{Velocity structure}

\begin{figure}
\centering
\includegraphics[width=1.0\linewidth]{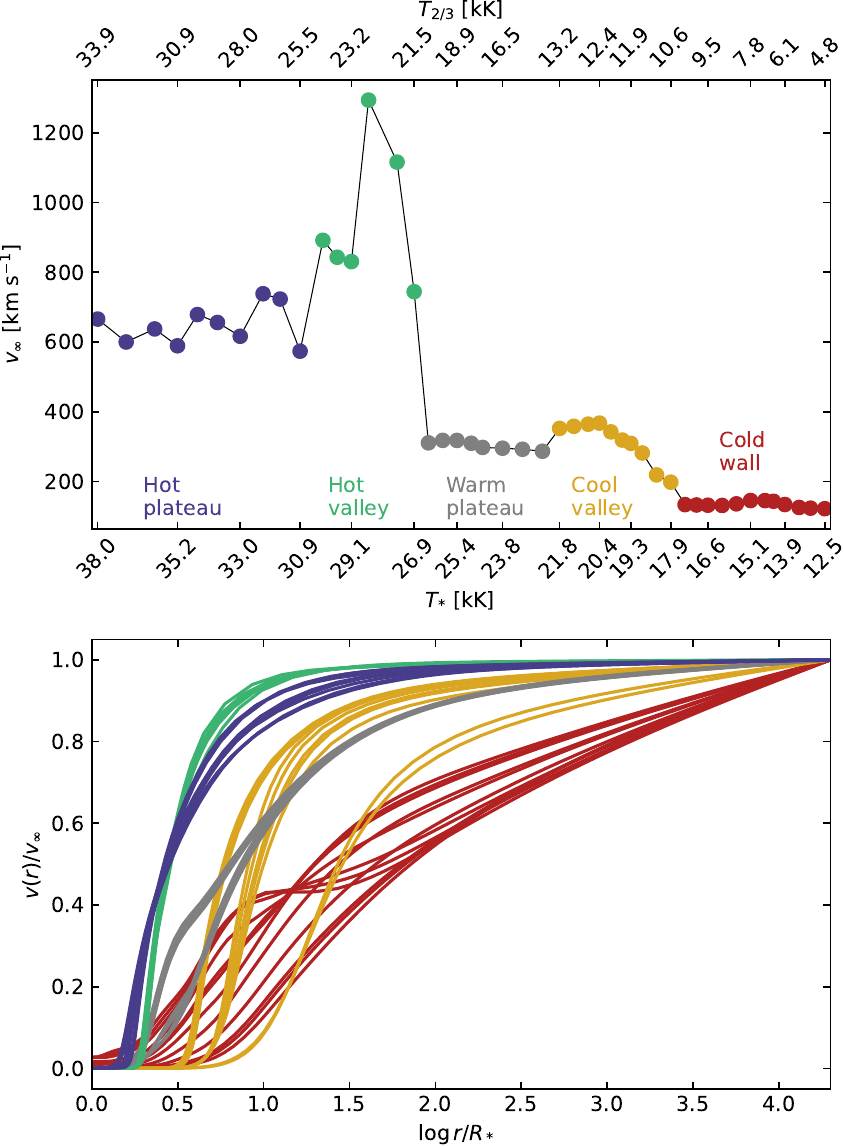}
  \caption{
    Terminal velocities (upper panel) and velocity structures (lower panel) of the models along the $T_\ast$ sequence. Each wind-solution domain is color-coded following the scheme of Fig.~\ref{fig:airy-dense-Rt}.
  }
    \label{fig:v_struc_shape}
\end{figure}

The velocity structures in the $T_*$ sequence reflect the different dense and rarefied solutions. The upper panel of Fig.~\ref{fig:v_struc_shape} shows that the trends in $\varv(r)$ trace the five different solution regimes presented above, namely models in the valleys see an increase in $\varv_\infty$ compared to models in the plateaus. The differences become even clearer in the shapes of $\varv(r)$, which show major structural differences for the different solutions (lower panel of Fig.\,\ref{fig:v_struc_shape}). None of the solutions is sufficiently described by a $\beta$-velocity law, especially in the wind-launching region. However, when we fit the solutions of the hot dense-wind plateau (OHG regime, blue curves in Fig.\,\ref{fig:v_struc_shape}) with a single $\beta$-law, we obtained solutions resembling the outer wind part with $\beta$-values between $1.8$ and $2.0$, explaining why higher $\beta$-values are often favored in traditional analyses of B super- and hypergiants.

The hot dense-wind plateau has $\varv_\infty$$\sim$700~km~s$^{-1}$ and a relatively sharp velocity gradient. In the hot rarefied-wind valley (BSG-like regime, green curves), this increases to $\sim$1000~km~s$^{-1}$ and an even more extreme gradient at the critical point. When we fit this regime with $\beta$-laws, we obtained slightly higher values of $\beta \sim 2.3$.
From the 1-BiSJ into the warm dense-wind plateau (gray curves), $\varv_\infty$ sharply drops to $\sim$300~km\,s$^{-1}$. In parallel, the shapes noticeably switch from a shallower velocity increase and a reduction of the maximum velocity gradient. The corresponding best $\beta$-values increase to $\sim$4 in this regime, although the representation of the shape in the outer wind clearly deviates from a $\beta$-law.

Within the cool, rarefied-wind valley (yellow curves), the solutions switch back to a steeper gradient (like in the hot valley previously), but commence farther out in the wind (i.e., larger $R_\mathrm{crit}$). The $\varv_\infty$, however, does not increase much relative to the warm plateau, while the best $\beta$ increases further to values between $8$ and $20$. After the 2-BiSJ, in the cold dense-wind wall, $\varv_\infty$ is rather constant despite a rather sharp increase in the mass-loss rates. This is a consequence of the support by gas and turbulent pressure in the outer wind (see Fig.~\ref{fig:v_struc_sound_turb}). As the pressure term drops only with $1/r$, it eventually becomes the dominant term in the outer wind and causes up to $\sim$50\% of the obtained $\varv_\infty$ value. This regime can no longer be meaningfully described by a $\beta$-law. Like in the warm plateau, the differences in $v(r)$ are strong, with the coolest models at $\sim$10$R_\star$ showing a velocity plateau due to the decreasing effect of \ion{Fe}{ii} (lowest panel in Fig.\,\ref{fig:acc-struct}).

\begin{figure}
\centering
\includegraphics[width=1.0\linewidth]{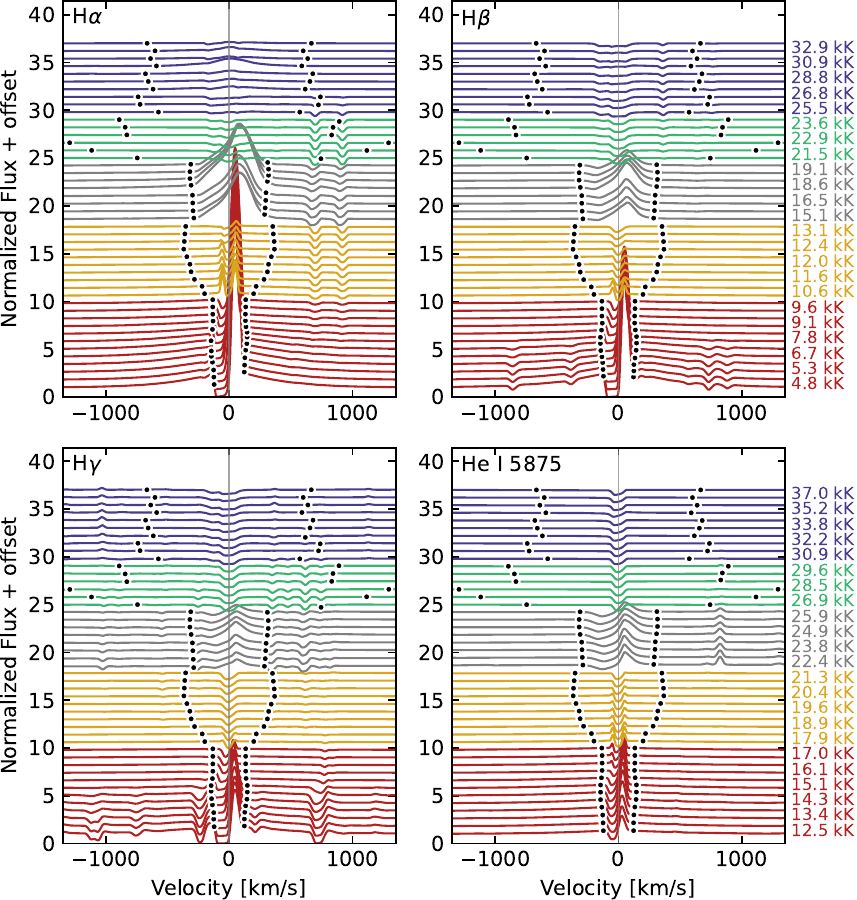}
  \caption{
    Hydrogen and helium profiles in velocity space compared to the terminal velocity (black dots). The color of each spectrum marks the wind-solution regime to which the model belongs, following the same color scheme as Fig.\,\ref{fig:v_struc_shape}. The temperatures rightward in the upper panels indicate $T_{2/3}$, and rightward in the lower panel, they indicate $T_\star$.
  }
    \label{fig:H-He-lines}
\end{figure}

The different wind regimes and velocity structures also leave their imprint on the resulting synthetic spectra. In Fig.\,\ref{fig:H-He-lines} we plot the normalized spectra around the first three Balmer lines as well as \ion{He}{i}\,5875\,\AA. In addition, we mark the locations of $\varv_\infty$ leftward and rightward of the central line wavelength. In the hottest regime as well as in the rarefied regimes, the lines do not reflect $\varv_\infty$, even when parts are in emission. However, for the warm plateau below the 1-BiSJ, at least H$\alpha$ traces $\varv_\infty$ when the full emission line width is considered. Below the 2-BiSJ, the P\,Cygni troughs of all Balmer lines trace the terminal velocity well, while for \ion{He}{i}\,5875\,\AA\, this is only true for the coolest models, in which He is entirely recombined in the outer atmosphere and can produce the blue absorption.

\section{Discussion}
\label{sec:discussion}

\subsection{Terminal and escape velocity ratio}

From conservation of energy, we expect that the terminal velocity of radiation-driven winds directly depends on $\varv_{\text{esc}, \Gamma_\mathrm{e}}$
at the stellar surface. For line-driven winds within the mCAK formalism,
\begin{equation}
\varv_\infty/\varv_\mathrm{esc, \Gamma_\mathrm{e}} \sim [\alpha/(1-\alpha)]^{1/2}\mathrm{,}
\end{equation}
where $\alpha$ is the parameter that condenses the blend of optically thin and thick sets of lines \citep[see, e.g.][]{Lamers+1995}. This means that changes in the total line opacities as a result of temperature changes, for example, will affect how $\varv_\infty$ and $\varv_\mathrm{esc, \Gamma_\mathrm{e}}$ scale. Using literature values and spectral type calibrations as well as $\Gamma_\text{e}$ estimates based on evolutionary tracks, \citet{Lamers+1995} found two jumps in the \vinfvesc ratio in the OBA supergiant regime. These were later interpreted as a consequence of changes in the dominant Fe ion \citep{Vink+1999}. More recent studies contested the sharpness of these jumps \citep[e.g.,][]{Crowther+2006, Markova_Puls2008}.

\begin{figure}
\centering
\includegraphics[width=1.0\linewidth]{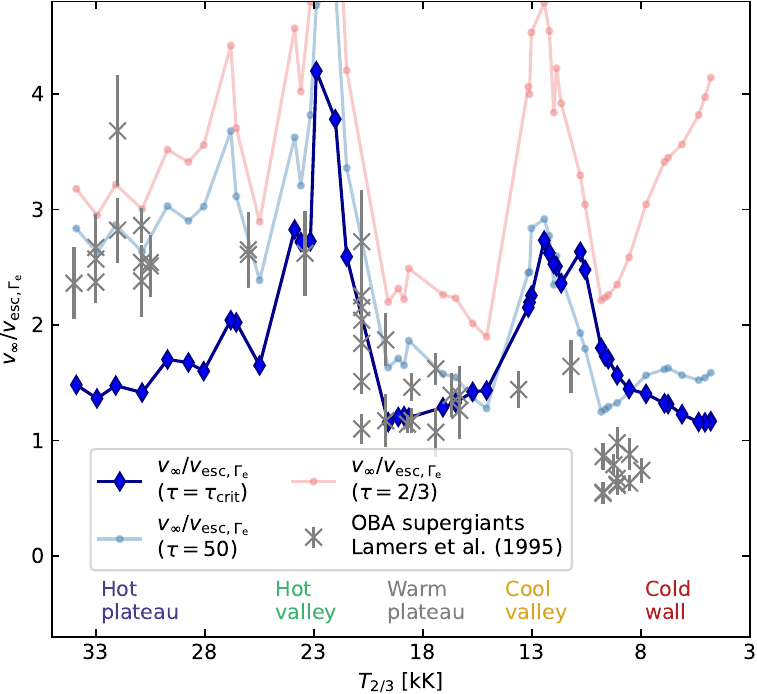}
  \caption{
Behavior of the $\varv_\infty/\varv_\mathrm{esc, \Gamma_e}$ relation. The gray crosses indicate the data of \cite{Lamers+1995} for OBA supergiants. The blue curve with diamonds indicates the values for $\varv_\mathrm{esc, \Gamma_e}$ applied to the critical point. The pink curve indicates the ratio at $\tau = 2/3$, and the thin light blue curve indicates the ratio at $\tau =50$, i.e., the inner boundary. 
  }
    \label{fig:vinfvesc}
\end{figure}

With our \texttt{PoWR$^\textsc{hd}$} model sequence, we tested the \vinfvesc-ratio and compare it to that of \citet{Lamers+1995} in Fig.\,\ref{fig:vinfvesc}.
For stars with extended atmospheres, the choice of $r$ in Eq.\,\eqref{eq:vescg} (where $\Gamma_\text{e}$ has to be evaluated as well) is not trivial, and we thus plot results for the inner boundary radius (corresponding to $\tau = 50$), $R_{2/3}$, and the critical radius $R_\text{crit}$. As the latter essentially reflects the wind onset, we considered this as the most coherent choice.
For models on the hot side of the classical 1-BiSJ, OBHGs have a lower \vinfvesc\ than OBSGs. This is mainly the result of the lower $\varv_\infty$ values of the hypergiants, which are lower by at least a factor of two than those of comparable supergiants (i.e., similar temperatures). Hence, although $\Gamma_\text{e}$ is higher for hypergiants, the change to $\varv_\mathrm{esc, \Gamma_\mathrm{e}}$ resulting from the $\sqrt{1-\Gamma_\text{e}}$ dependence in Eq.\,\eqref{eq:vescg} is much smaller.

In the hot rarefied valley, the terminal velocities become significantly higher because the lower mass-loss rates are lower. In this regime, some models even exceed the ratio found by \citet{Lamers+1995} for BSGs (but in agreement with some results reported in \citet{Markova_Puls2008}), which is a combined effect of the higher $\varv_\infty$ and a lower $\varv_\mathrm{esc, \Gamma_\mathrm{e}}$ caused by a larger $R_\text{crit}$.
In the 1-BiSJ regime, where \ion{Fe}{IV} recombines to \ion{Fe}{III}, the \vinfvesc ratio drops sharply to values only slightly above unity, which agrees well with the results of \citet{Lamers+1995}. While the $\varv_\infty$ values are very similar to those in the hot valley, the drop in the \vinfvesc ratio is mainly caused by the deeper wind-launching, resulting in lower $R_\text{crit}$ values (i.e., higher $\varv_\mathrm{esc, \Gamma_\mathrm{e}}$).

In the cool valley, before the 2-BiSJ, the ratios increase again. While a glimpse of this might be seen in the results from \citet{Lamers+1995}, as well as in some of the recent \texttt{METUJE} calculations by \citet{Krticka+2025}, and possibly also in the empirical determinations by \citet{Markova_Puls2008}, we obtained a much more pronounced effect, again for the same reasons as in the first valley. Crossing the 2-BiSJ, where the \ion{Fe}{III}$\rightarrow$\ion{Fe}{II} switch occurs, then coincides with another drop in \vinfvesc, which is also reflected in the data from \cite{Lamers+1995}, albeit with an offset. However, since the turbulent pressure term affects $\varv_\infty$ considerably in this regime, as discussed above, we might overestimate the ratio by up to a factor of two, which would bring our data in line with those of \cite{Lamers+1995}, despite their rough temperature assumptions from spectral types for the B and A stars. 

Our \vinfvesc\ in the A-star regime are also larger than the values obtained from the so-called $\delta$-slow solutions by \cite{Cure+2011} and empirical values listed therein. Again, we attribute this to the effect of the pressure term in the outer wind, which consists of a gas pressure and a turbulent pressure component. In the current framework, both components are treated together as an isotropic pressure. The turbulent pressure is further described with a constant $\varv_\text{turb}$. While the value is clearly motivated by observations \citep[e.g.,][]{Searle+2008}, simulations of the multidimensional time-dependent hot star atmosphere in the WR- and O-type star domain from \citet{Moens+2022} and \citet{Debnath+2024} suggested that radiatively driven turbulence is no longer isotropic in the supersonic wind. While a more detailed treatment of radiatively driven turbulence requires a major update of the 1D hydrodynamic framework as well as calibration with multidimensional models, it is clear that solutions with a leading $1/r$-acceleration in the outer wind have to be critically reviewed. For the coolest models, Fig.\,\ref{fig:v_struc_sound_turb} clearly shows that the turbulent velocity exceeds the sound speed, meaning that the outermost wind solution is dominated by $\varv_\text{turb}$, similar to the models in \citet{Sabhahit+2026}. These results, as well as the ongoing discussions about turbulent pressure as a driver for cool massive star outflows \citep[e.g.,][]{Stothers2003,Klochkova2019,Kee+2021}, underline that turbulent pressure is an important ingredient at the cool end of the massive star regime, requiring the development of a more coherent framework addressing the interplay between radiative acceleration and radiatively driven turbulence in launching mass outflows.

Another caveat when discussing terminal velocities in hydrodynamically consistent modeling is the choice of wind clumping, which can affect the calculated $\varv_\infty$ by enhancing opacities. We fixed $f_\text{cl} = 3$ throughout the whole sequence to a value motivated for the BSG regime by the line-deshadowing instability simulations from \citet{Driessen+2019}. As discussed in their work and as also empirically supported, the clumping in the supersonic region in early-B and O supergiants (i.e., on the hot side of the 1-BiSJ) is expected to be much larger ($f_\mathrm{cl}$$\sim$$20$) than in later-type BSGs. Therefore, it is plausible that $\varv_\infty$ is systematically underestimated for the hottest models of our sequence (i.e., in the hot plateau). This, however,  partially contradicts our finding of a very good match between the UV P\,Cygni profiles of the OHG GS~Mus, but agrees with the mismatch for HD151804, which has a much higher $\varv_\infty$.

\subsection{The $\varv_\infty$--$T_\mathrm{eff}$ relation}

\begin{figure}
\centering
\includegraphics[width=1.0\linewidth]{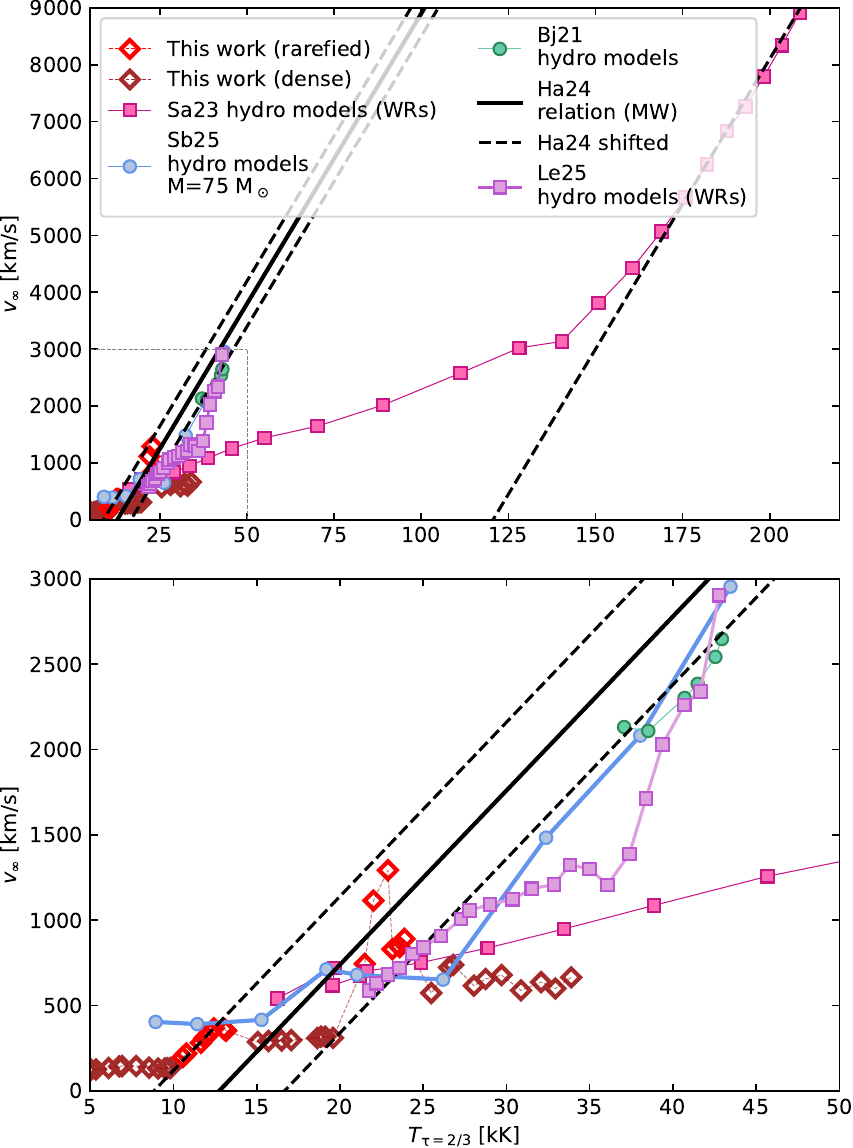}
  \caption{
  Relation of $\varv_\infty$--$T_\mathrm{eff}$ for different studies that made hydrodynamically consistent sequences of models in which $T_\star$ varied. The black lines indicate the empirical relation by \cite{Hawcroft+2024} (the solid line represents the actual relation, and the dashed line indicates the relation shifted arbitrarily). The pink and violet squares indicate the models by \cite{Sander+2023} and \cite{Lefever+2025}, respectively. The blue circle indicates the models of \cite{Sabhahit+2026} with $\Gamma_e \sim 0.35$, and the green circle indicates the models of \cite{Bjorklund+2021} with similar $\Gamma_e$. The red and brown diamonds indicate our sequence of models and show the rarefied and dense solutions, respectively. The lower panel is a zoom-in of the region within the dotted rectangle in the upper panel.
  }
    \label{fig:hydro-vinf-teff}
\end{figure}

From $\varv_\infty \propto \varv_{\text{esc},\Gamma}$ and the Stephan-Boltzmann equation, we derive the relation 
\begin{equation}
    \varv_\infty = \mathcal{C}(\kappa[X_\mathrm{He},Z,\rho,f_\mathrm{cl}]) \,\,\,\, L^{1/4} T_\mathrm{eff} \sqrt{\frac{1-\Gamma_\mathrm{e}}{\Gamma_\mathrm{e}}}\text{,}
\end{equation}
where $\mathcal{C}$ is a constant that depends on the opacity $\kappa$, which depends on the chemical composition ($X_\mathrm{He}$, $Z$), density $\rho$, and clumping factor $f_\mathrm{cl}$. This relation explains the empirical scaling of $\varv_\infty$ with $T_\mathrm{eff}$. In Fig.\,\ref{fig:hydro-vinf-teff} we plot our resulting values using  $T_\mathrm{eff} = T_{2/3}$. The different dense and rarefied solutions leave clear imprints as they show different slopes in the $\varv_\infty$--$T_\mathrm{eff}$ relation with sharp breaks precisely where the wind becomes denser (i.e., in the dense-wind plateaus). While there is more scatter for the hotter models, the rarefied solutions generally match the empirically obtained slope from \cite{Hawcroft+2024} (but shifted in $T_\text{eff}$ for the cooler valley), while in the dense-wind regimes, the relation becomes much flatter.

In Fig.\,\ref{fig:hydro-vinf-teff} we also plot recent other temperature sequences of hydrodynamically consistent models (with otherwise homogeneous parameters) in the OB and WR regime. Other studies with $\Gamma_\mathrm{e} \sim 0.3$ tend to agree well with the slope of the relation reported by \citeauthor{Hawcroft+2024}. Likewise, when their winds become thicker, the slopes become flatter. While the slopes in the dense-wind regime disagree with those of the WR models from \citet{Sander+2023}, from which we show the $12.9\,M_\odot$ sequence with $\Gamma_\text{e} \sim 0.32$, the thin-wind regime also matches the (shifted) empirical relation by \cite{Hawcroft+2024}, indicating that it might be generally valid throughout the whole line-driven wind regime down to $\sim$10~kK, independent of $\Gamma_\text{e}$, as long as the winds do not become dense. Even the near-Eddington models from \citet{Lefever+2025} agree mostly with the \cite{Hawcroft+2024} slope, similar to our rarefied-wind solutions with a departure toward the trend from the WR sequence for $T_{2/3} > 28\,$kK. 

Below $\sim$10~kK, that is, past the 2-BiSJ, the stars approach the edge of the line-driven wind regime, and their mass loss might be generated and sustained by other sources, such as thermal and turbulent pressures and pulsations \citep[e.g,][]{Glatzel-Kraus2024,Glatzel-Kiriakidis2013,Fadayev2010}.
As discussed above, most of the outer wind in our models in this temperature regime is sustained by gas and turbulent pressure. Turbulent pressure also plays a role in launching, although radiative acceleration remains the major contributor there, as shown in Fig.\,\ref{fig:grand-acc-mdot} and Fig.\,\ref{fig:acc-struct}. For the coolest models, the dependence on the turbulence is strongest as a viable wind (i.e., $\Gamma_\mathrm{wind} > 1$) is not expected to occur without this term, and different values of $\varv_\mathrm{turb}$ will change $\varv_\infty$. This is also shown in Fig.\,\ref{fig:hydro-vinf-teff}, where the model sequence of \citet{Sabhahit+2026} with similar $T_{2/3}$ has higher $\varv_\infty$ values precisely due to the much higher $\varv_\mathrm{turb} = 70$\,km\,s$^{-1}$.

\subsection{Interpretation of the mass-loss trend and comparison with existing recipes}

As shown in Sect.~\ref{sec:results}, the $\dot{M}$ trend we obtained with plateaus and valleys is benchmarked by the resemblance of the corresponding synthetic spectra to several hypergiants and LBVs in the OBA star domain, and it is thus expected to capture the atmospheric physics in this high-$\Gamma_\mathrm{e}$ regime.
The jumps in $\dot{M}$ (1-BiSJ and 2-BiSJ) are clearly caused by the recombination of Fe, as predicted by \cite{Vink+1999}. 
In contrast, no clear jumps are observed in BSGs, despite changes in Fe ionization at the base of the wind \citep[e.g.,][]{Bernini-Peron+2023,Bernini-Peron+2024,deBurgos+2024,Alkousa+2025}. This clearly indicates a difference that to first order originates in the proximity to the Eddington limit as the BSGs are in a lower $\Gamma_\mathrm{e}$ regime. Together with the recent calculations in \citet{Sabhahit+2026} probing different $\Gamma_\text{e}$-sequences (although with fixed $L$), the complexity of the derived trend and the bistability of the solutions obtained near the jumps suggest a general bimodality of dense and rarefied-wind solutions. For sufficient conditions, a dense-wind solution with high mass-loss rates can be reached where $\dot{M}$ even seems to increase with decreasing temperature. However, when these conditions are not met, the solutions switch to the rarefied branch, where mass-loss rates tend to decline with decreasing temperature. When we adopt this generalized paradigm, the spectra of our findings reveal that hypergiants and some late-type LBVs reside in a transition domain in which the atmospheric conditions can switch between rarefied and dense solutions, depending on the temperature.

While the two solutions do not correspond to the transition to optically thick winds (i.e., $\tau_\text{crit} > 1$), the transition is still triggered by the availability or absence of sufficient opacity (plus turbulent pressure), thus marking a striking similarity with the regime differences found for stripped He-burning stars in \citet{SanderVink2020} and \citet{Sander+2023}. Comparing our findings with the different sequences from \citet{Sabhahit+2026} and \citet{Bjorklund+2023}, we suggest that depending on $\Gamma_\text{e}$, there might be no switch between dense and rarefied regimes because for high $\Gamma_\text{e}$ (and sufficient metallicity, see below), the whole sequence can be in the dense-wind regime where continuum and geometrical effects will dominate. In contrast, for low $\Gamma_\text{e}$, all solutions fall onto the rarefied branch, as likely occurred in the models from \citet{Bjorklund+2023} and \citet{Krticka+2024,Krticka+2025}. However, to confirm this hypothesis, a systematic exploration of the precise conditions is required to isolate the contributions of supplementary effects, such as clumping and turbulent pressure. 

Comparing our results with existing mass-loss recipes, we found that the recipe from \cite{Krticka+2024} applied to our model properties (considering $T_{2/3}$) matches $\dot{M}$ in our models in the rarefied-wind regime very well. Their recipe successfully reproduces the typical mass-loss rates determined for cooler BSGs \citep[e.g.,][]{Bernini-Peron+2023, Bernini-Peron+2024, Verhamme+2024}. The recipe by \citeauthor{Krticka+2024} also predicts a mild increase in the warm plateau region, albeit with a temperature offset. The recipe by \cite{Bjorklund+2023} (applied to $T_{2/3}$) in the hot plateau and relatively cool valley matches very well, but underestimates $\dot{M}$ for cooler temperatures.
When we compared our results to those by \citet[applied to $T_{2/3}$]{Vink+2001}, we found a decent match to our sequence up to the cool valley, with a systematic offset of 0.8\,dex. Interestingly, when we applied \citeauthor{Vink+2001} to our sequence, considering $T_\star$ and shifting the results down by this offset, we obtained an excellent match with our results for the hot plateau, hot valley, and warm plateau (including the 1-BiSJ).
All the recipes fail catastrophically to explain the behavior past the 2-BiSJ. At the coolest temperatures, the unscaled \citeauthor{Vink+2001} rates again match the $\dot{M}$ values we derived.

\begin{figure}
\centering
\includegraphics[width=1.0\linewidth]{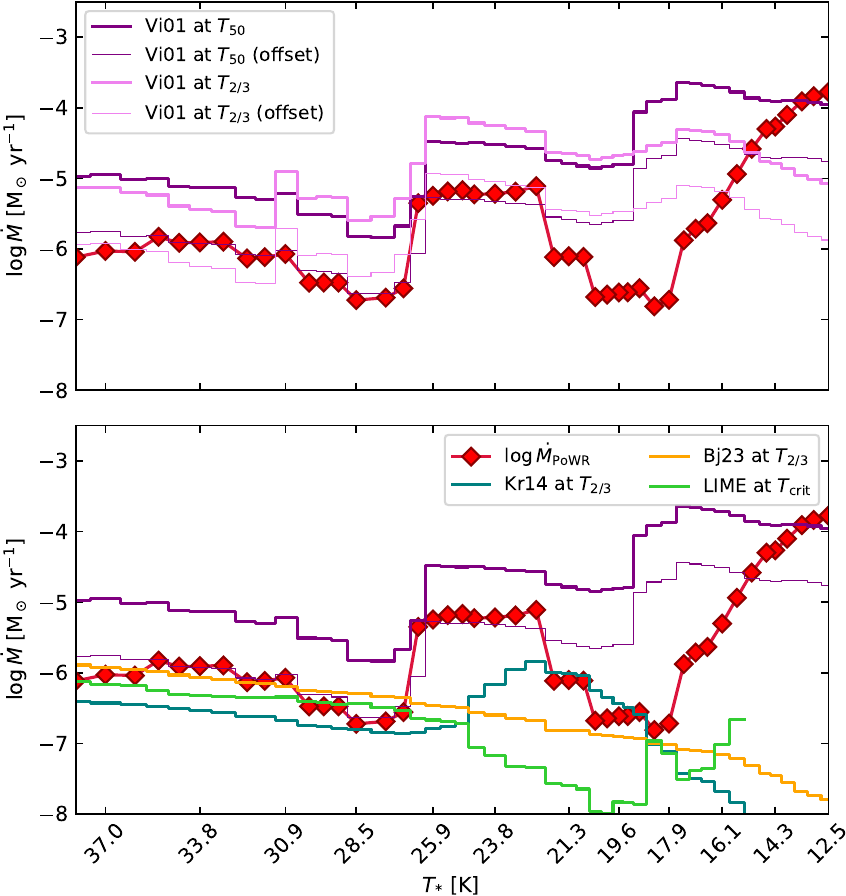}
  \caption{
Comparison between the output mass-loss rates of the $T_\star$ sequence (red curve) and different recipes applied to the stellar properties of each model. The thick purple and fuchsia lines indicate the mass-loss rates by \cite{Vink+2001}, applied to $T_\star$ and $T_{2/3}$, respectively. The thin lines of the same colors indicate the Vink rates with an offset of 0.8 dex. The yellow curve indicates the rates of \cite{Bjorklund+2023}, and the teal line indicates the rates of \cite{Krticka+2024}, both applied to $T_{2/3}$. The green line indicates the rates computed using the tool LIME \citep[\url{https://lime.ster.kuleuven.be/},][]{Sundqvist+2025}
  }
    \label{fig:mdot-comparison}
\end{figure}

\subsection{The effect of metallicty on the cool models}

\begin{figure}
\centering
\includegraphics[width=1.0\linewidth]{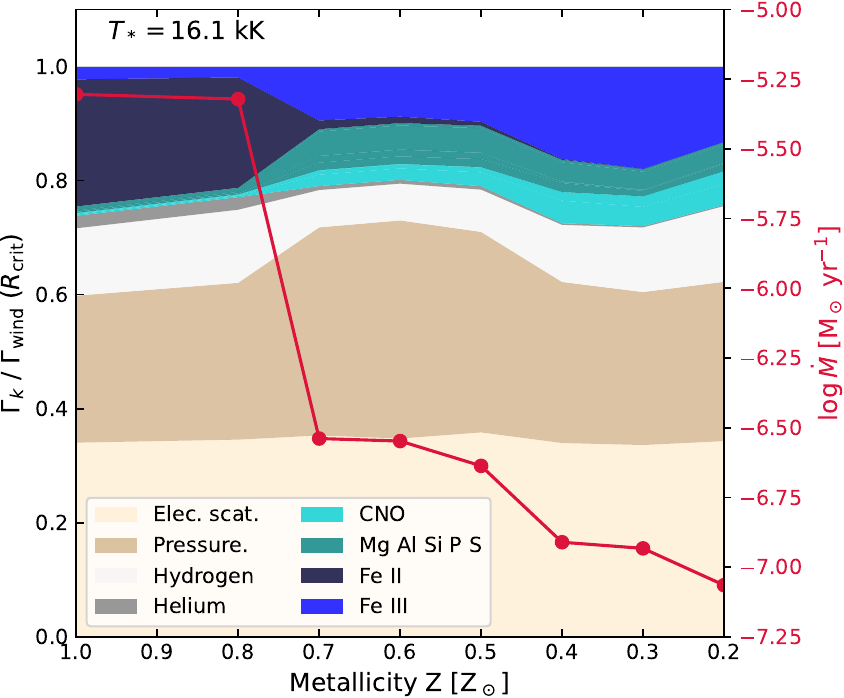}
  \caption{
Contribution to the radiative acceleration from different sources, elements, and ions (stacked colored areas) and mass-loss rates (red curve) according to metallicity for a fixed temperature of $T_\star = 16.1$\,kK.
  }
    \label{fig:ion_acc_mdot_z}
\end{figure}

One of the most striking results of our study is the increase in $\dot{M}$ toward the coolest temperatures. As we discussed above, this is considerably different from what has been obtained in recent modeling efforts in the B supergiant regime. To investigate in particular the increase due to the ionization change from \ion{Fe}{iii} to \ion{Fe}{ii}, we also created auxiliary shorter sequences for a fixed $T_\star = 16.1\,$kK in this regime, where we changed the He composition and metallicity. In Fig.\,\ref{fig:ion_acc_mdot_z} we present the results, where we varied the metallicity of the model, and hence, also the Fe content. It is noticeable that between 0.7 and 0.6 $Z_\odot$, $\dot{M}$ drops by 1.5 dex, causing the models to switch back to a rarefied solution.

This additional sequence confirms that the line opacities are important for the switch between the two regimes, boosted in particular by the recombination of \ion{Fe}{iii} to \ion{Fe}{ii}. The \ion{Fe}{ii} opacities cause the mass-loss rates to increase, thus increasing the wind density. In turn, this causes H to recombine and further reduce the wind temperature. When $Z$ is sufficiently low, the wind is thinner, which allows more photons to ionize \ion{Fe}{ii} into \ion{Fe}{iii}. As noticeable in Fig.\,\ref{fig:ion_acc_mdot_z}, the relative role of the pressure term tends to increase while the shift in the Fe ion dominance occurs, which we also found in the main model sequence (Fig.\,\ref{fig:ion_acc_mdot}). A similar $Z$ dependence of a solution regime change seems to occur for the \ion{Fe}{iv} to \ion{Fe}{iii} transition in the models in \citet{Lefever+2025}, where a sharp transition in the $\dot{M}$ trend occurs between $0.8\,Z_\odot$ and $0.7\,Z_\odot$, while a smoother behavior is found at the highest and lowest $Z$. This would indicate that for the lowest metallicity, all solutions are on the rarefied branch, while at the highest $Z$, the solutions might all be on the dense branch.

\begin{figure}
\centering
\includegraphics[width=1\linewidth]{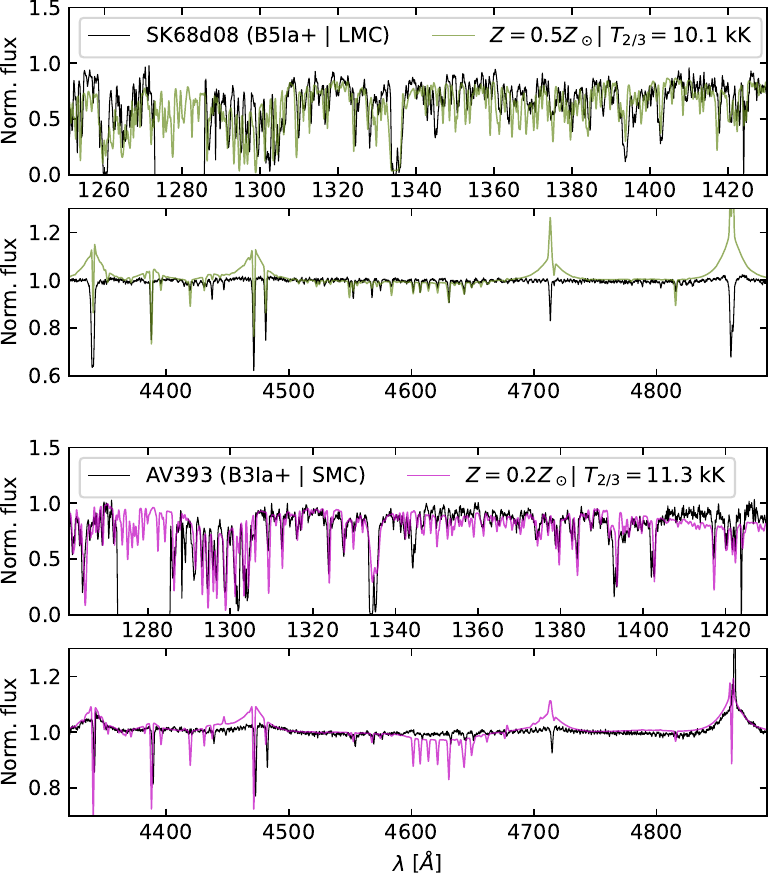}
  \caption{
  Nontailored comparison between the $0.5\,Z_\mathrm{\odot}$ and $0.2\,Z_\mathrm{\odot}$ models (olive and violet lines, respectively). The observed spectra of the BHGs SK\,$-68^{\circ}$08 (B5Ia+, in the LMC) and AzV~393 (B3Ia+, in the SMC) are represented by black lines. 
  }
    \label{fig:LMC-SMC-stars}
\end{figure}

Similar to the solar metallicity sequence, we also found that existing late-type BHGs in the LMC and SMC resemble the normalized spectra resulting from our lower-$Z$ models. Namely, for $Z = 0.2\,Z_\mathrm{\odot}$, we found a good match between the late BHGs AzV~393 (B3Ia+, SMC) and the synthetic UV and optical spectra (see Fig.~\ref{fig:LMC-SMC-stars}). For $Z = 0.5\,Z_\mathrm{\odot}$, we also found signatures comparable to the late BHG Sk~-68~8 (B5Ia+, LMC).
The $L$, $M$, and $X_\mathrm{He}$ values of these stars are in particular very similar to our metallicity model sequence \citep[][]{Verhamme+2024, Alkousa+2026}, such that the spectral mismatches, especially a systematic excess in the electron scattering wings, might be an indication of a different clumping stratification, which in case of hydrodynamically consistent models affect the output atmosphere structure and spectrum.

\subsection{Comparisons with yellow hypergiants}

\begin{figure}
\centering
\includegraphics[width=1\linewidth]{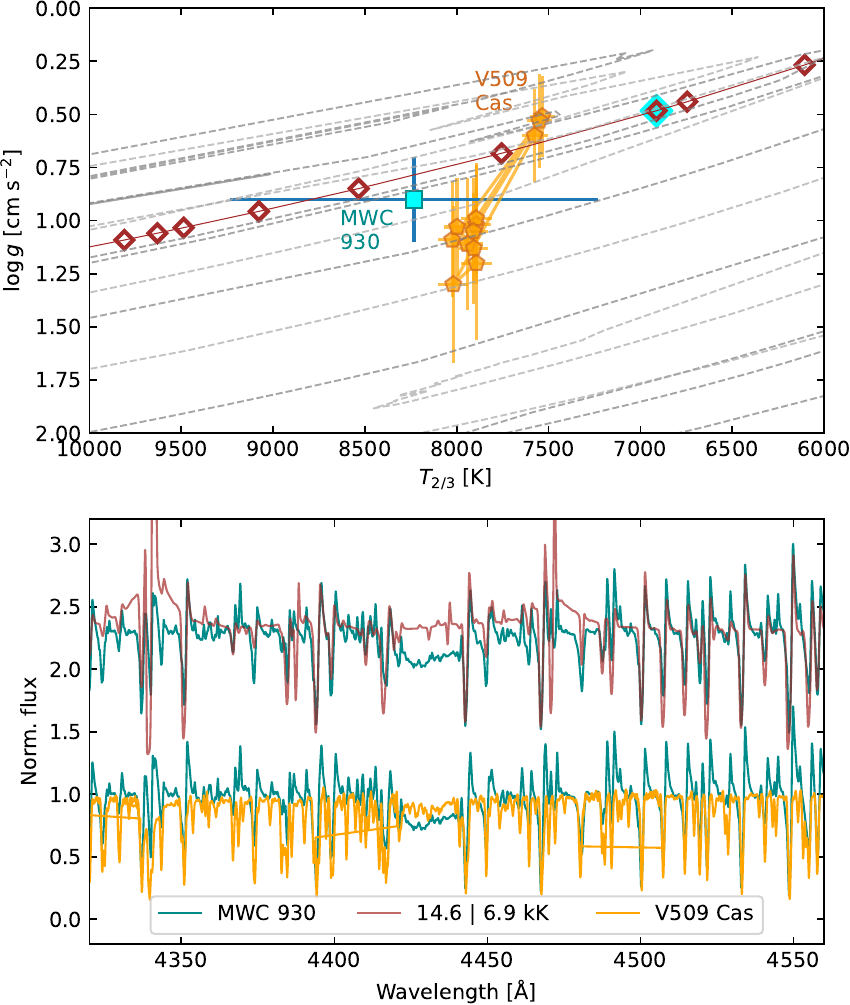}
  \caption{
  \textit{Upper panel:} Comparison between our model sequence (brown diamonds), MWC~930 at its maximum \citep[][cyan square]{Miroshnichenko+2014}, and the YHG V509~Cas after 2000 \citep[][golden pentagons]{Nieuwenhuijzen+2012} in the Kiel diagram. The cyan diamond marks the model that resembles MWC~930 best. \textit{Lower panel:} Comparison between the respective spectra. The brown plot shows our model that is most similar to MWC~930, and the golden plot shows the spectrum of V509~Cas in 2012. The spectrum of the LBV is plotted underneath the two as a dark cyan line for comparison. The thin dotted line in the background indicates the tracks of \cite{Ekstroem+2012}.
  }
    \label{fig:Kiel-spec}
\end{figure}

When comparing the position in the HRD and sHRD of our models against the positions of some known YHGs, we noted that their empirically obtained $T_\mathrm{eff}$ \citep[e.g.,][]{Nieuwenhuijzen+deJager2000,Nieuwenhuijzen+2012,Kasikov+2024} and our coolest models overlap. 
Taking V509\,Cas as a representative example, the effective temperature after the year 2000 remains consistently between $\sim$7.5 and $\sim$8.2 kK \citep{Kasikov+2024}. 
Considering the $T_\mathrm{eff}$ and $\log g$ values from \cite{Nieuwenhuijzen+2012}, we found that our model sequence crosses the variability track of this star in the sHRD parameter space essentially orthogonally (see Fig.\,\ref{fig:Kiel-spec}).

In the lower panel of Fig.\,\ref{fig:Kiel-spec}, we compare the spectrum of the YHG V509\,Cas to that of the LBV MWC~930 during its maximum, which is closely resembled by our model with $T_\star = 14.6\,$kK. It is evident that the LBV-type spectrum essentially resembles the YHG spectrum, but is enhanced by wind features. Notably, both stars have $T_\mathrm{2/3} \ll 10$\,kK, but in the LBV case, this is due to the photosphere being formed beyond the critical point, thus not representing the $T_\text{eff}$ of the (quasi-)hydrostatic atmosphere. Our model resembling MWC~930 has $T_\mathrm{2/3} \sim 7$\,kK, while V509\,Cas is formally hotter. 

The formal distinction between supergiants and hypergiants in the yellow regime is different from that in the BA-star regime and is not based on the same spectroscopic criterion as was used for OBA hypergiants \citep[e.g.,][]{deJager+1998, Kasikov+2026}. Therefore, LBVs such as MWC~930 at its maximum might very well be what more extreme BHGs with dense winds are to BSGs or even BHGs in the rarefied regime. While our model sequence does not extend to cooler temperatures where \ion{Fe}{i} becomes more important, the trend of finding rarefied solutions when the Fe ionization stage at $r \sim R_\mathrm{crit}$ is shifting suggests that another rarefied-wind regime might appear at even cooler $T_\star$, which might then produce observed YHG spectra such as that of V509\,Cas.

\section{Conclusions}
\label{sec:conclu}

We conducted a thorough analysis with hydrodynamically consistent atmosphere models of a parameter space representative of OBA hypergiants and LBVs at solar metallicity to investigate their mass-loss rates and wind velocity structure in this crucial snapshot in the evolution of high-mass stars.
We highlight the following conclusions:
\begin{itemize}
    \item We found two solution branches, which we described as dense and rarefied. In the explored temperature sequence, the models mainly settled on the dense solution, but were unable to maintain this in certain parameter regions. These regions appear as valleys in the $\dot{M}(T_\text{eff})$ diagram. The dense solutions are associated with higher wind optical depths, but do not require values exceeding unity. Instead, the rarefied solutions in this work can be described by $R_\text{t} > 200\,R_\odot$, although more work is needed to verify whether this value is universal.
    \item The normalized spectra synthesized from our models resemble the main spectral appearance of various OBA hypergiants and LBVs. Spectral resemblance is found for the dense and rarefied solutions. Exemplary calculations indicate that these findings might even hold at lower metallicity.
    \item We found that the contributions of the electron scattering and thermal pressure components cause $\sim$50\% of the force at the wind onset (i.e., critical point).
    \item Our model sequence results agree with recent near-Eddington \texttt{PoWR$^\textsc{hd}$} calculations employing different luminosities and masses than in our study. \citet{Sabhahit+2026} also reported valley-like features in some of their $\dot{M}(T_\star)$ sequences, in particular, those matching our $\Gamma_\text{e}$-regime and farther away from the Eddington limit. The higher-$\Gamma_\text{e}$ models calculated by \citet{Lefever+2025} at solar metallicity do not display the pronounced valleys, but seem to be in a transition regime between our solutions and even higher $\Gamma_\text{e}$-sequences in \citet{Sabhahit+2026}. However, the $\dot{M}$ jumps occurring at lower metallicities in \citet{Lefever+2025} show clear similarities to the jumps obtained in this work.
    \item In contrast to modeling efforts for OB stars, we found a general increase in the mass-loss rate with decreasing temperature. We suggest that this behavior might be exclusive to the dense branch. The solutions in the rarefied branch are notably supported by turbulent pressure, and our absolute results thus differ from calculations such as those by \citet{Bjorklund+2023} and \citet{Krticka+2024}, although each of the recipes matches partially in some temperature regimes. Down to $T_\star \sim 22\,$kK, our derived mass-loss pattern matches the prediction by \citet{Vink+2001} well when it is shifted downward by $0.8\,$dex.
    \item The solution switched from rarefied to dense in our sequence when Fe recombined, such that \ion{Fe}{III} or \ion{Fe}{II} can provide significant opacity at the critical point. This sharp increase in the mass-loss rate coincides with the traditional bistability jumps from \citet{Vink+1999}. We also found an actual bistability of solutions at these jumps. 
    \item For LBVs (P~Cygni supergiants) with dense winds on the cool side of the second jump, the terminal velocity is well traced by the H lines, but not necessarily by the He\,I lines.
    \item We found that the empirical linear relation by \cite{Hawcroft+2024} of the wind terminal velocity with the stellar photospheric temperature seems to be valid for all rarefied-winded stars that are confirmed to be in the line-driven wind regime ($T_\mathrm{eff} \gtrsim 10$\,kK), that is, it is valid from late BHGs to WRs. However, the different classes of stars seem to have a different offset.
    \item We provide updated luminosities for galactic OBA hypergiants and LBVs based on corrected Gaia DR3 parallaxes when available or applicable. The updated luminosity distribution causes the population of LBVs and hypergiants (assuming properties publicly available in the literature) to agree better with the LBV instability strip. However, careful and thorough analysis is necessary to confirm whether this agreement between theory and empirical findings actually holds.
    
\end{itemize}

\begin{acknowledgements}
      The authors thank A.\ Kasikov for inspiring discussions and sharing some forthcoming results before publication.
      The authors thank S.\ Sim\'{o}n-Diaz and A.\ de\ Burgos for providing spectroscopic data from the IACOB project as well as inspiring discussions. The authors also thank J.\ Josiek, R.\ R.\ Lefever, G.\ Gonz{\'a}lez-Tor{\`a}, E.\ Sch{\"o}sser, and D.\ Pauli for helpful discussions during this manuscript's preparation. Moreover, the authors thank the anonymous referee for the constructive comments and suggestions, which were fundamental to improve the quality of this manuscript.
      MBP and AACS are supported by the Deutsche Forschungsgemeinschaft (DFG, German Research Foundation) in the form of an Emmy Noether Research Group – Project-ID 445674056 (SA4064/1-1, PI Sander). VR acknowledges financial support by the Federal Ministry for Economic Affairs and Energy (BMWE) via the Deutsches Zentrum f\"ur Luft- und Raumfahrt (DLR) grant 50 OR 2509 (PI Sander). MBP is a member of the International Max Planck Research School for Astronomy and Cosmic Physics at the University of Heidelberg (IMPRS-HD). This project was co-funded by the European Union (Project 101183150 - OCEANS). Based on observations collected at the European Southern Observatory under ESO programme(s). This work presents results from the European Space Agency (ESA) space mission Gaia. Gaia data are being processed by the Gaia Data Processing and Analysis Consortium (DPAC). Funding for the DPAC is provided by national institutions, in particular the institutions participating in the Gaia MultiLateral Agreement (MLA).
      FN acknowledges support by PID2022-137779OB-C41 funded by MCIN/AEI/10.13039/501100011033 by ``ERDF A way of making Europe'' and grant MAD4SPACE, TEC-2024/TEC-182 from Comunidad de Madrid (Spain).
\end{acknowledgements}

\bibliographystyle{aa}
\bibliography{biblio.bib}

\begin{appendix}

\section{Atomic data}
\label{sec:atomic}

\begin{table}[]
\caption{\label{tab:atomic} Ions, levels, and transitions (lines) considered in the models}
\begin{tabular}{lccc|lccr}
\hline\hline
Elem. & ion & levels & lines & Elem. & ion & levels & lines \\
\hline
H  & I   & 30 & 435  &    Mg & I   & 1  & 0    \\
H  & II  & 1  & 0    &   Mg & II  & 32 & 496  \\
He & I   & 45 & 990  &  Mg & III & 43 & 903  \\
He & II  & 30 & 435  &  Mg & IV  & 17 & 136   \\
He & III & 1  & 0    &     Mg & V   & 20 & 190  \\
N  & I   & 10 & 45   &  Si & I   & 20 & 190    \\
N  & II  & 38 & 703  &  Si & II  & 20 & 190     \\
N  & III & 85 & 3570 &  Si & III & 24 & 276     \\
N  & IV  & 38 & 703  &  Si & IV  & 55 & 1485    \\
N  & V   & 20 & 190  &  Si & V   & 52 & 1326   \\
N  & VI  & 14 & 91   &  P  & IV  & 12 & 66     \\
C  & I   & 15 & 105  &  P  & V   & 11 & 55      \\
C  & II  & 32 & 496  &  P  & VI  & 1  & 0       \\
C  & III & 40 & 780  &  Al & I   & 1  & 0       \\
C  & IV  & 25 & 300  &  Al & II  & 10 & 45       \\
C  & V   & 29 & 406  &  Al & III & 10 & 45   \\
C  & VI  & 1  & 0    &  Al & IV  & 1  & 0   \\
O  & I   & 13 & 78   &  Fe & I    & 13 & 40 \\
O  & II  & 37 & 666  &  Fe & II   & 14 & 48 \\
O  & III & 33 & 528  &  Fe & III  & 13 & 40 \\
O  & IV  & 29 & 406  &  Fe & IV   & 18 & 77 \\
O  & V   & 54 & 1431 &  Fe & V    & 22 & 107 \\
O  & VI  & 35 & 595  &  Fe & VI   & 29 & 194 \\
O  & VII & 15 & 105  &  Fe & VII  & 19 & 87 \\
S  & II  & 32 & 496  &  Fe & VIII & 14 & 49 \\
S  & III & 23 & 253  &  Fe & IX   & 15 & 56 \\
S  & IV  & 25 & 300  &  Fe & X    & 1  & 0 \\
S  & V   & 20 & 190  &   &    &      &         \\
S  & VI  & 22 & 231  &  &    &      &       \\
\hline
\end{tabular}
\end{table}

In Table\,\ref{tab:atomic}, we list the elements and ions with the number of levels and line transitions considered in our atmosphere models. For Fe -- which in \texttt{PoWR} refers to all iron-group elements conflated --, the levels and line numbers refer to superlevels and ``superlines'' \citep[see][for more details]{Graefener+2002}. The $\varv_\mathrm{dop}$ parameter, which controls the resolution of the frequency-dependent cross section of all the elements \citep{Graefener+2002} is set to the same value of the turbulent motion -- i.e., $\varv_\mathrm{dop} = 20$~km~s$^{-1}$.

\section{Wind structure in different regimes}
\label{sec:wind-acc-struc}

\begin{figure}
    \centering
    \includegraphics[width=1\linewidth]{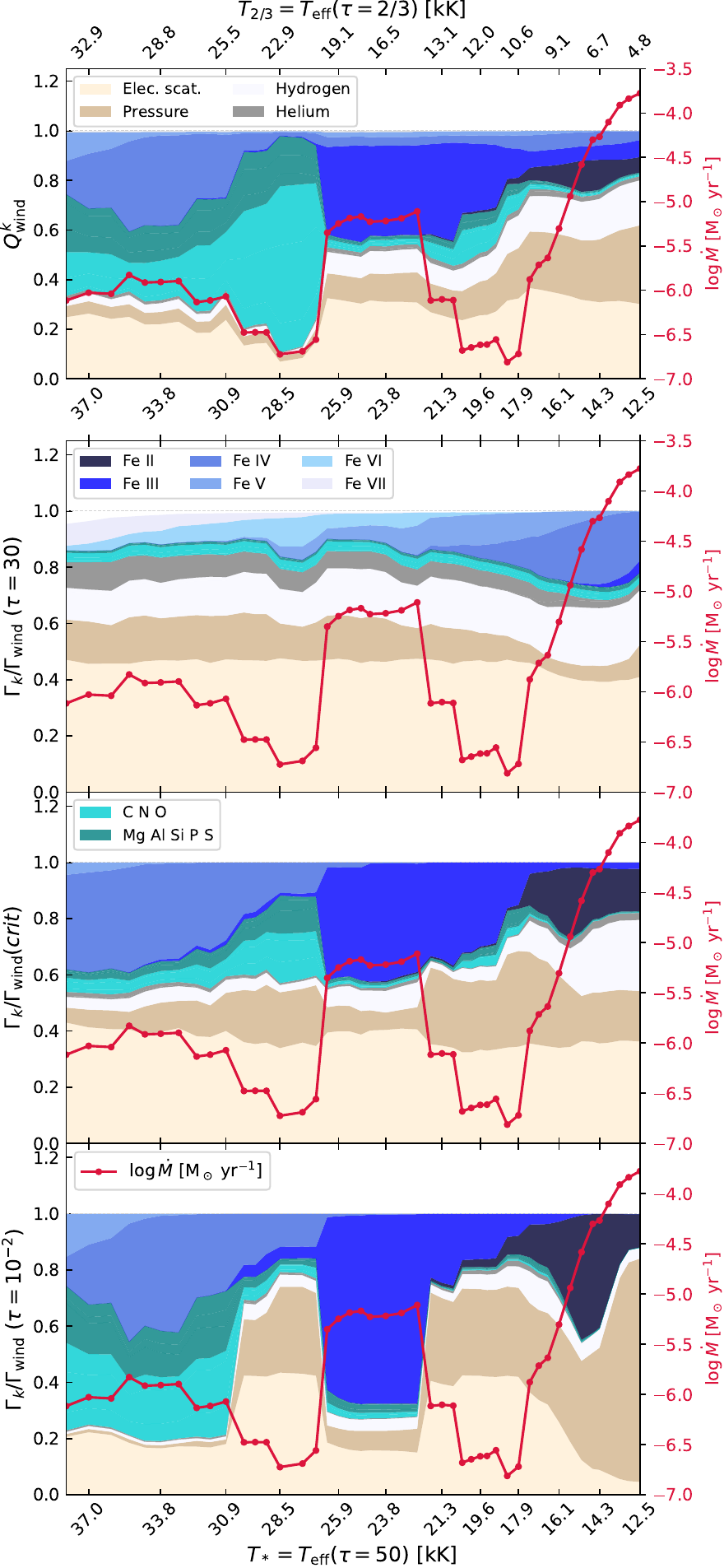}
    \caption{\textit{Upper panel:} Contribution of different processes, elements, and ions to the wind work ratio $Q_\mathrm{wind}$. The other panels indicate each contribution to the wind driving at specific depths -- namely, at $\tau = 30$ (near the inner boundary), $\tau = \tau_\mathrm{crit}$ (where the wind is launched, like in Fig.\,\ref{fig:ion_acc_mdot}), and $\tau = 0.01$ (far out in the wind). The color code is the same of Fig.\,\ref{fig:ion_acc_mdot}.}
    \label{fig:grand-acc-mdot}
\end{figure}

In Fig.\,\ref{fig:ion_acc_mdot} we compare $\dot{M}$ with the contribution of the main processes, elements, and ions to the wind acceleration at the critical point for each value of $T_\star$. In Fig.\,\ref{fig:grand-acc-mdot} we provide an extension of this figure showing the respective main contributions to the wind driving at different points in the atmosphere. Namely, at $\tau = 20$, near the inner boundary ($\tau = 50$), probing the forces in the inner atmosphere, (ii) at the critical point, where the wind is effectively launched, and (iii) at $\tau = 0.01$, which probes conditions far out in the wind.

\begin{figure}
\centering
\includegraphics[width=1.0\linewidth]{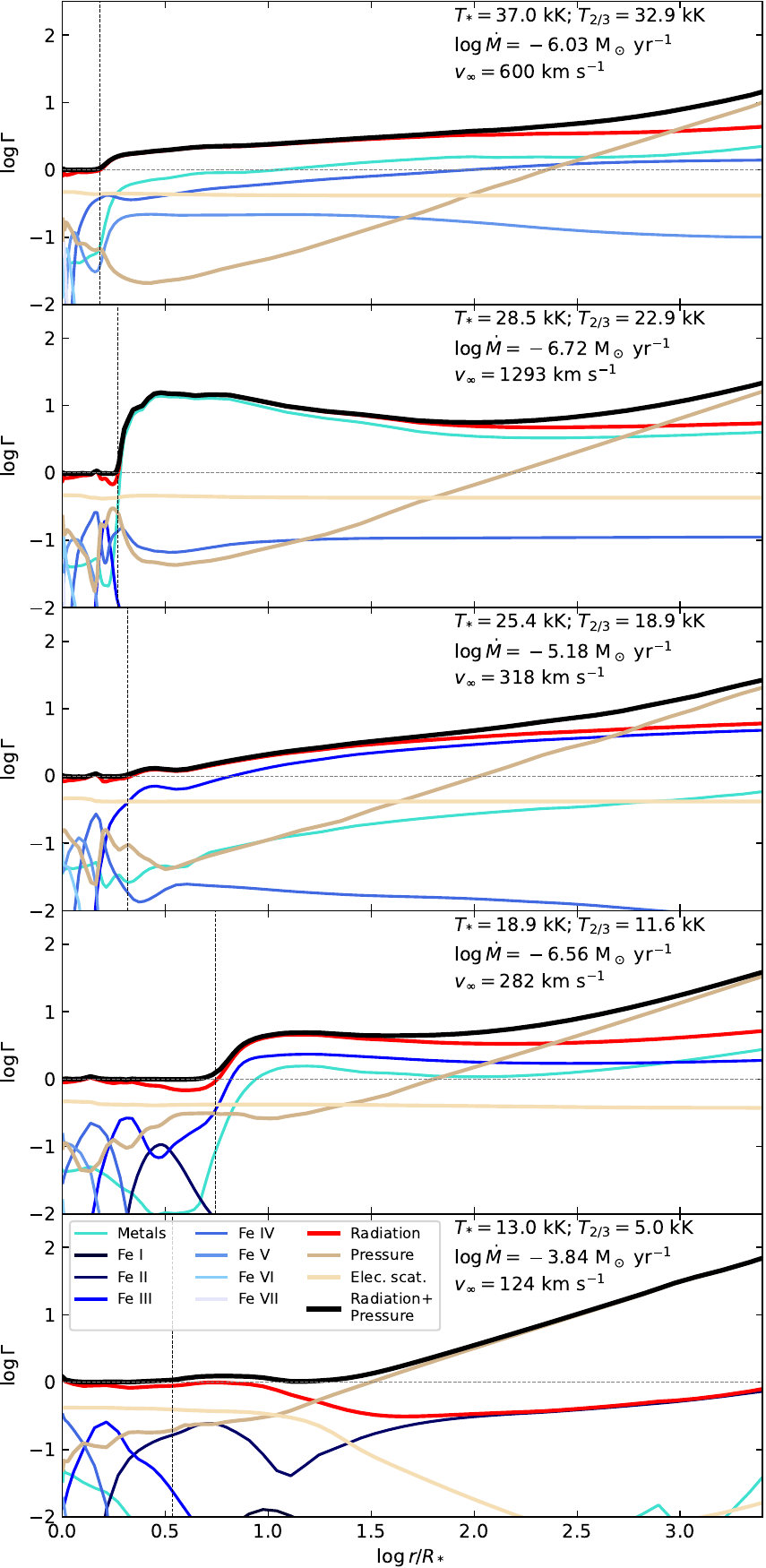}   
  \caption{
Acceleration structure and main contributions to the driving of the winds for models in different wind-solutions domains.
  }
    \label{fig:acc-struct}
\end{figure}

\begin{figure}
\centering
\includegraphics[width=1.0\linewidth]{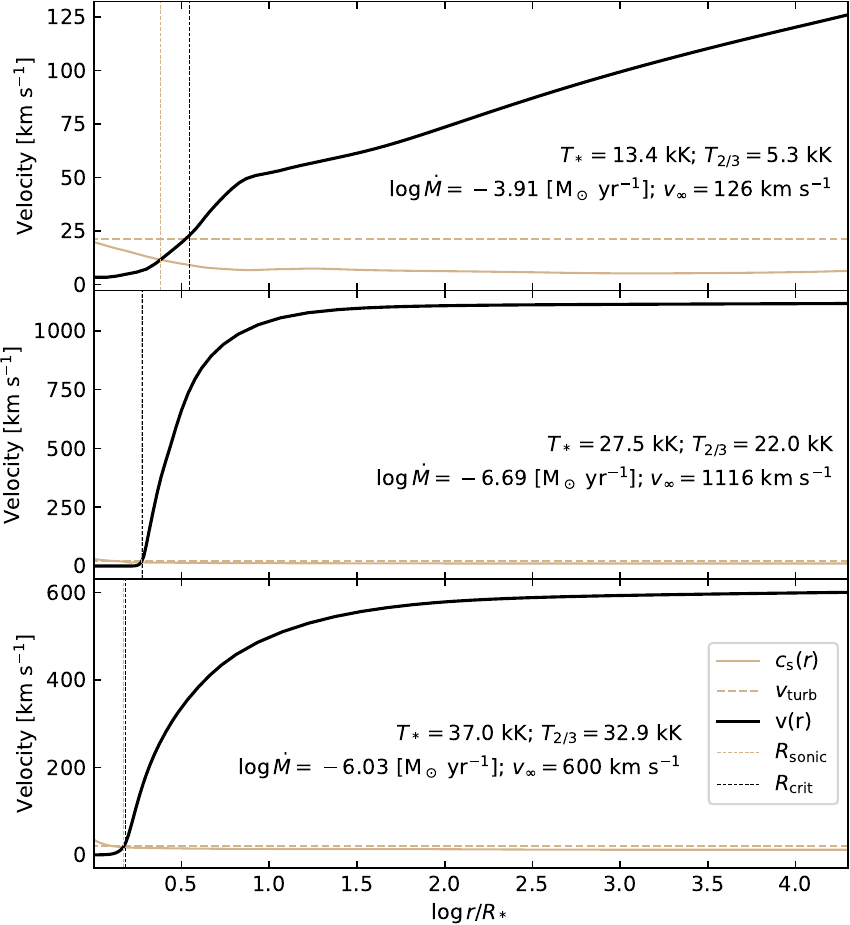}
  \caption{
    Wind velocity structure (black line) for models in the cold wall, hot valley, and hot plateau wind-solution regions. The beige line indicates the sonic speed structure ($c_\mathrm{s}$) while the dotted beige line indicate the turbulent velocity $\varv_\mathrm{turb} = 20$~km\,s$^{-1}$. The thin vertical lines indicate the sonic radius ($R_\mathrm{sonic}$, beige) and critical radius ($R_\mathrm{crit}$, black).
  }
    \label{fig:v_struc_sound_turb}
\end{figure}

A more detailed view on the acceleration structure for representative models at each regime described in Sect.\,\ref{sec:results} is given in Fig.\,\ref{fig:acc-struct}. In Fig.\,\ref{fig:v_struc_sound_turb} we illustrate the velocity structure of models in different regimes. One can see that the turbulent velocity is equal to or larger than the sound speed in the wind, indicating that the turbulent pressure dominates over the gas thermal pressure. This is particularly important for models in the cold wall.

\section{Bistability of the solutions at the jumps}
\label{app:bistable}

When computing hydrodynamically consistent models, the complex interplay between radiation field, species populations, and velocity field does not necessarily converge to unique solutions. 
Such complexity is due to the nonlinearity of the equations being solved, coupled in space and frequency. Therefore, the radiative acceleration structure, obtained via the integration of the momentum equation inward and outward from the critical point, depends on conditions elsewhere in the atmosphere.

This means that, for instance, different initial physical conditions can yield different results (with them all satisfying the criteria for numerical convergence) or even fail to reach convergence in case the initial ``guess'' is far from a valid solution. Given the high computational cost of each model, exhaustively mapping these degeneracies is beyond the scope of this work. However, as a sanity check of the obtained trend, we compute models with different initial conditions in distinct regions of our $T_\mathrm{eff}$ sequence.

\begin{figure}
    \centering
    \includegraphics[width=1\linewidth]{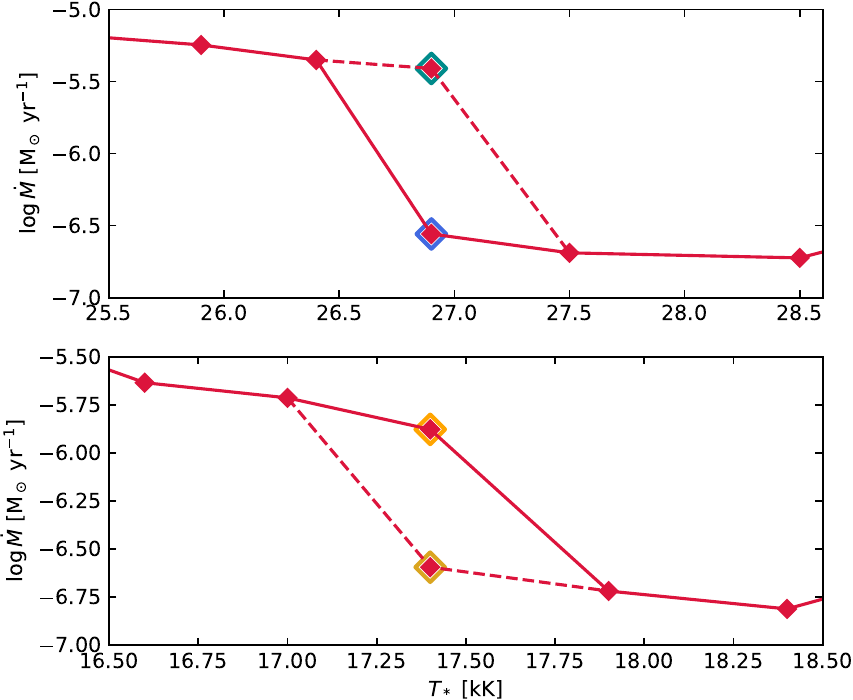}
    \caption{Demonstration of the bistability of solutions in the 1-BiSJ (upper panel) and 2-BiSJ (lower panel). In either case, we find models with the same $T_\ast$ yielding different results based on their initial conditions. }
    \label{fig:hysteresis}
\end{figure}

This is particularly important when operating in the transition between the rarefied and dense solutions. In Fig.~\ref{fig:hysteresis}, we show that at the edge of the 1-BiSJ and 2-BiSJ, the same $T_\star$ can produce different values of $\dot{M}$ when the initial conditions are a model on the rarefied or on the dense branch. This also manifests in the spectral appearance of the models' synthetic spectra, which also match different observed stars.

\begin{figure}
    \centering
    \includegraphics[width=1\linewidth]{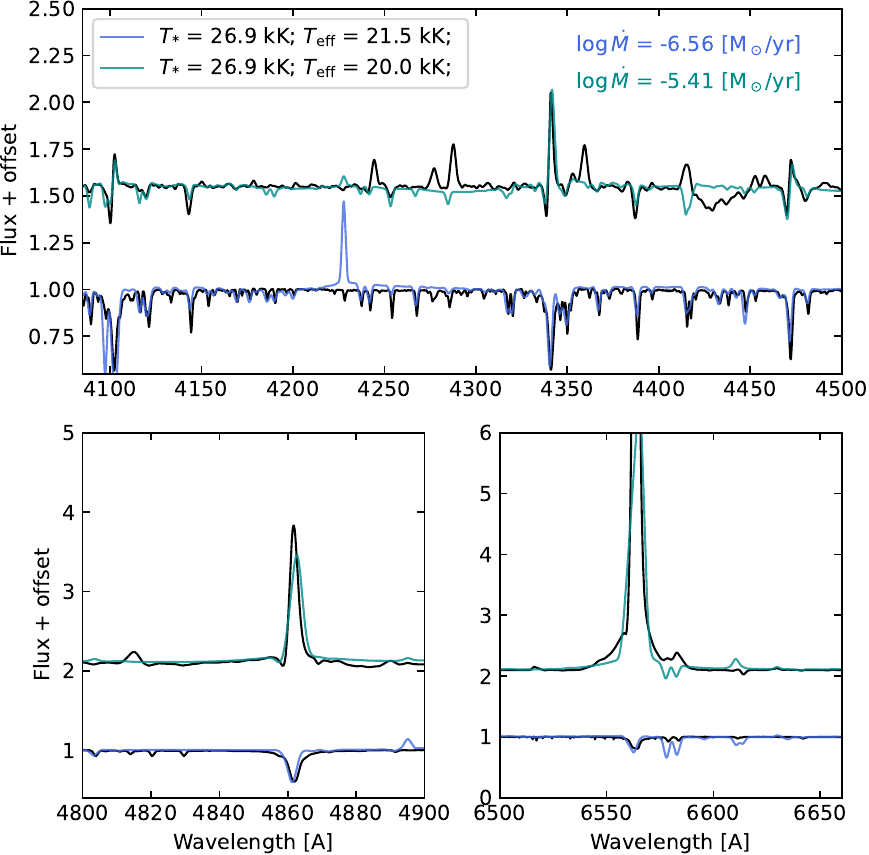}
    \caption{Manifestation of the bistability of solutions in the model spectrum for the 1-BiSJ in our sequence.}
    \label{fig:degen-269}
\end{figure}

\begin{figure}
    \centering
    \includegraphics[width=1\linewidth]{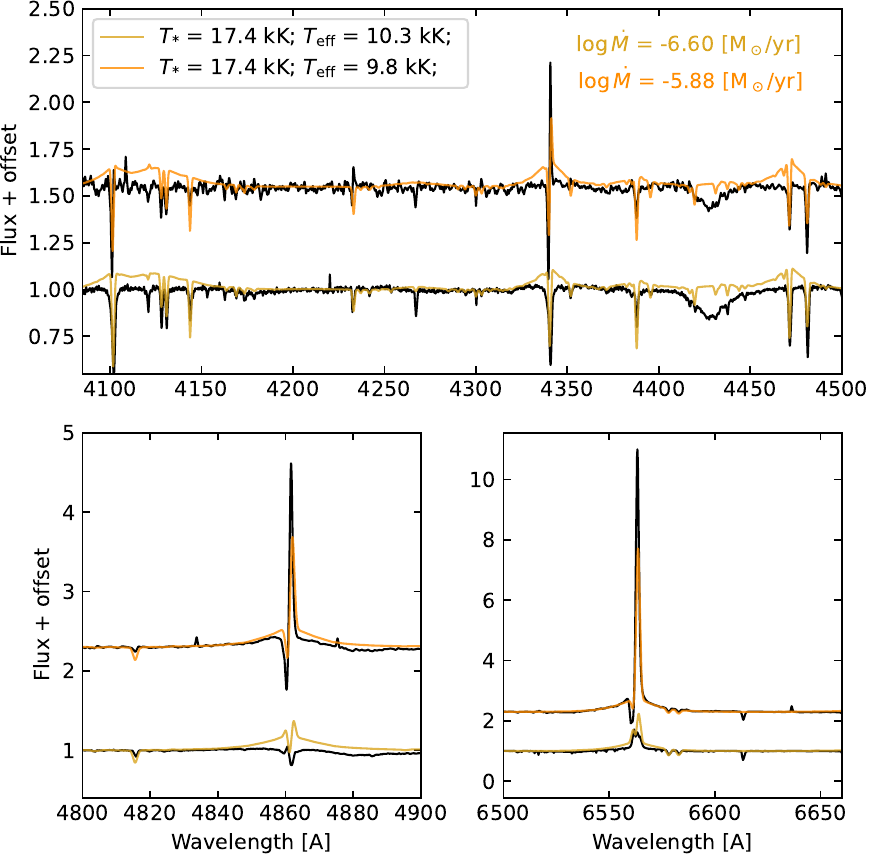}
    \caption{Analogous to Fig.\,\ref{fig:degen-269} but for the 2-BiSJ.}
    \label{fig:degen-174}
\end{figure}

The strong dependence on initial conditions prohibits us from pinpointing one-to-one relations between temperatures
(or other parameters) and wind properties. This exercise of computing different models with different initial conditions can also provide a certain interval of confidence in our $\dot{M}(T_\mathrm{eff})$ behavior and the associated transition between regimes. However, it indicates that stars with similar properties may exist in different regimes -- which recycles the original discussions on bistability by \cite{Pauldrach_Puls1990}.

\subsection{Mechanical wind luminosity}
\label{sec:Lmech}

\begin{figure}
\centering
\includegraphics[width=1.0\linewidth]{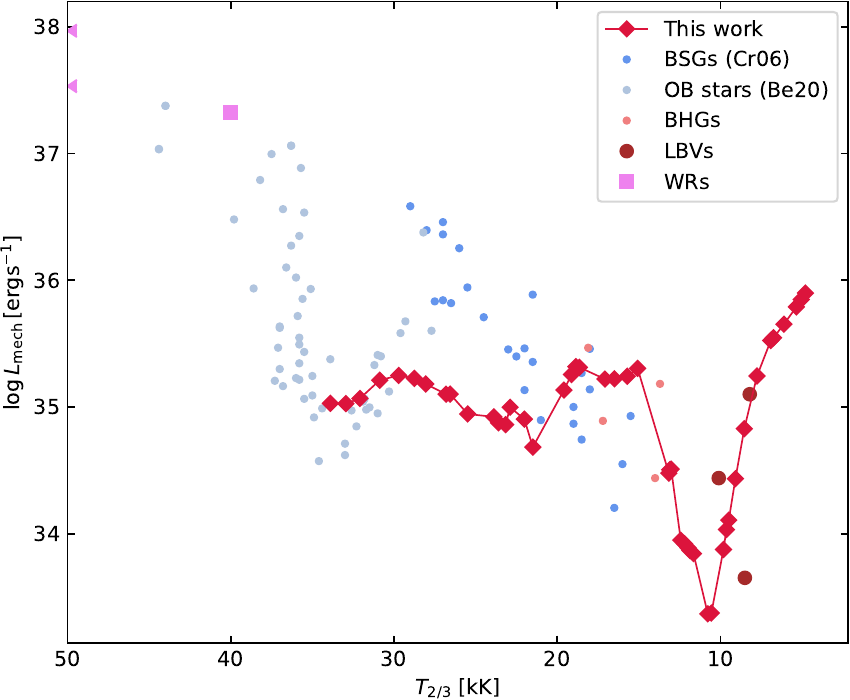}
  \caption{
Mechanical luminosity of the model sequence compared to literature data from \citet{Crowther+2006}, \citet{Berlanas+2020} -- light blue circles --, and WR $L_\text{mech}$ estimates in \citet{Vieu+2024} -- pink squares and arrows, which indicate WRs outside the plotting area.
  }
    \label{fig:Lmech}
\end{figure}

To investigate the impact of hypergiants on their environment, we calculate the mechanical wind luminosity $L_\mathrm{mech}:= 0.5\,\dot{M}\,\varv_\infty^{2}$. In general, up to the cool rarefied valley, we find $L_\mathrm{mech} \sim 10^{36}\,\mathrm{erg\,s^{-1}}$, comparable to late-O and B stars \citep[see, e.g.,][]{Vieu+2024}. In the valley, $L_\mathrm{mech}$ plunges due to the combination of very low $\dot{M}$ and $\varv_\infty$. Past the 2-BiSJ, the high mass-loss rates reached in the cold wall region invert the trend and cause $L_\mathrm{mech}$ to increase up to $\sim$$10^{36}\,\mathrm{erg\,s^{-1}}$. However, the low terminal velocities lead to generally lower $L_\text{mech}$ values, making hotter stars and in particular Wolf-Rayet stars significantly more important with respect to kinetic feedback unless much higher luminosities and thus mass-loss rates would be reached than mapped in our model sequence. Nonetheless, significant mass loss from cooler massive stars can affect cluster properties via mass loading \citep[e.g.,][]{Larkin+2025a,Larkin+2025b}.

\section{Luminosity of GS Mus (O9.7Ia+) and HD151804 (O8Iaf)}
\label{sec:OHG-lum}

\begin{figure}
    \centering
    \includegraphics[width=1\linewidth]{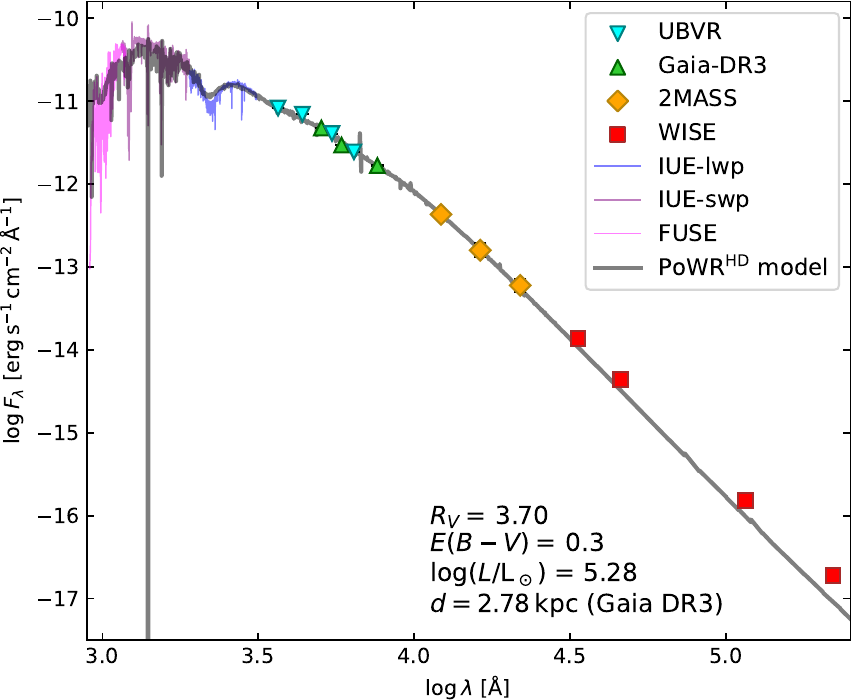}
    \caption{Spectral energy distribution of GS~Mus. The photometry used in were sourced from \cite{GaiaCollab+2023}, \cite{Paunzen+2022}, \cite{Zacharias+2005}, and \cite{Cutri+2013}.}
    \label{fig:GSMus-SED}
\end{figure}

\begin{figure}
    \centering
    \includegraphics[width=1\linewidth]{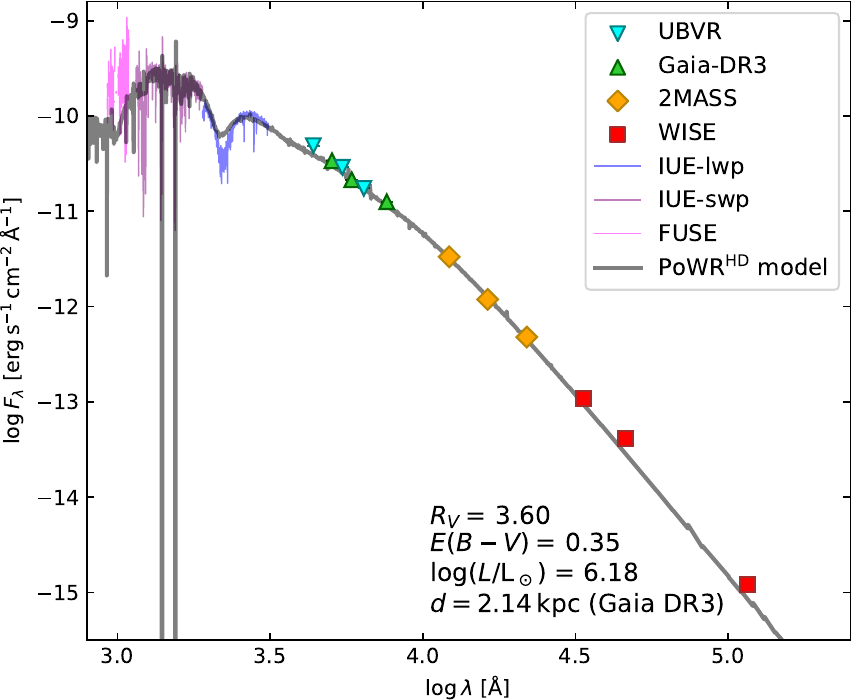}
    \caption{Spectral energy distribution of HD151804. The photometry is sourced from the same references as in Fig.~\ref{fig:GSMus-SED}.}
    \label{fig:HD151804-SED}
\end{figure}

In the literature, we were able to find spectroscopic properties of GS~Mus and HD151804 obtained via analysis of the normalized spectra \citep[e.g.,][]{Prinja+1990,Holgado+2020}.
However, we could not find luminosities based on SED fitting. For GS~Mus \cite{Lefever+2007} derived $\log L/\mathrm{L_\odot} = 5.75$ based on the $V$ magnitude, spectral type and luminosity class.
As our models yielded similar $T_{2/3}$ and $\log g$, we estimated their luminosities via fitting the optical and IR photometric fluxes and flux-calibrated far-UV -- see Fig.~\ref{fig:GSMus-SED} and Fig.~\ref{fig:HD151804-SED}. The reddening was estimated via fitting the near-UV 2175~$\mathrm{\AA}$ absorption bump applying the extinction law by \cite{Cardelli+1989}. As we do not perform a dedicated fitting, we consider a conservative uncertainty of 0.2~dex to the luminosity of each star.

\section{Model fundamental wind parameters}
\label{sec:model_prop}

In Tables \ref{tab:wind-fund-props} and \ref{tab:wind-fund-props-Z}, we disclose the fundamental atmospheric properties of the sequences of models discussed in this paper.

\begin{table*}[]
    \centering
    \caption{Atmosphere fundamental properties for each model in the $T_\star$ sequence.}
    \label{tab:wind-fund-props}
\begin{tabular}{lccccccccr}
\hline\hline
$T_\star$ & $R_\star$ & $\log\dot{M}$ & $T_{2/3}$ & $R_\mathrm{t}$ & $\log g$ & $\Gamma_\mathrm{e} (R_\mathrm{crit})$ & $R_{2/3}$ & $R_\mathrm{crit}$ & $R_\mathrm{sonic}$ \\

[kK] & [$R_\odot$] & [$\mathrm{M}_\odot\,\mathrm{yr}^{-1}$] & [kK] & [$R_\odot$] & [cm\,s$^{-2}$] & -- & [$R_\star$] & [$R_\star$] & [$R_\star$] \\ 
\hline
12.5 & 131.8 & -3.78 & 4.8 & 8.69 & -0.16 & 0.39 & 6.82 & 3.29 & 2.18 \\
13.0 & 121.9 & -3.84 & 5.0 & 8.89 & -0.06 & 0.39 & 6.63 & 3.41 & 2.30 \\
13.4 & 114.7 & -3.91 & 5.3 & 9.45 & 0.04 & 0.39 & 6.29 & 3.52 & 2.40 \\
13.9 & 106.6 & -4.10 & 6.1 & 12.31 & 0.27 & 0.39 & 5.18 & 3.73 & 2.58 \\
14.3 & 100.7 & -4.27 & 6.7 & 15.67 & 0.44 & 0.39 & 4.50 & 3.97 & 2.74 \\
14.6 & 96.6 & -4.30 & 6.9 & 16.00 & 0.48 & 0.39 & 4.46 & 4.07 & 2.80 \\
15.1 & 90.3 & -4.58 & 7.8 & 23.04 & 0.68 & 0.39 & 3.79 & 4.95 & 3.06 \\
15.6 & 84.6 & -4.94 & 8.5 & 35.70 & 0.85 & 0.38 & 3.34 & 6.43 & 3.67 \\
16.1 & 79.5 & -5.30 & 9.1 & 57.31 & 0.96 & 0.39 & 3.15 & 8.55 & 4.88 \\
16.6 & 74.8 & -5.64 & 9.5 & 89.88 & 1.03 & 0.39 & 3.06 & 10.59 & 6.06 \\
17.0 & 71.3 & -5.71 & 9.6 & 97.18 & 1.06 & 0.40 & 3.12 & 11.24 & 6.50 \\
17.4 & 68.0 & -5.88 & 9.8 & 119.82 & 1.09 & 0.40 & 3.15 & 12.50 & 7.28 \\
17.9 & 64.3 & -6.72 & 10.6 & 534.50 & 1.22 & 0.42 & 2.88 & 11.16 & 8.33 \\
18.4 & 60.8 & -6.81 & 10.8 & 624.17 & 1.25 & 0.42 & 2.92 & 10.90 & 8.33 \\
18.9 & 57.7 & -6.56 & 11.6 & 472.15 & 1.39 & 0.42 & 2.63 & 5.56 & 4.85 \\
19.3 & 55.3 & -6.61 & 11.9 & 526.10 & 1.42 & 0.42 & 2.65 & 5.42 & 4.79 \\
19.6 & 53.6 & -6.62 & 12.0 & 523.21 & 1.44 & 0.42 & 2.67 & 5.35 & 4.76 \\
20.0 & 51.5 & -6.65 & 12.2 & 551.28 & 1.47 & 0.42 & 2.69 & 5.22 & 4.70 \\
20.4 & 49.5 & -6.68 & 12.4 & 585.08 & 1.50 & 0.42 & 2.70 & 5.11 & 4.64 \\
20.8 & 47.6 & -6.11 & 13.0 & 233.92 & 1.58 & 0.42 & 2.55 & 3.68 & 3.46 \\
21.3 & 45.4 & -6.10 & 13.1 & 217.74 & 1.59 & 0.42 & 2.64 & 3.78 & 3.56 \\
21.8 & 43.3 & -6.12 & 13.2 & 209.33 & 1.60 & 0.42 & 2.74 & 3.93 & 3.71 \\
22.4 & 41.1 & -5.11 & 15.1 & 36.98 & 1.84 & 0.42 & 2.21 & 2.79 & 2.22 \\
23.1 & 38.6 & -5.19 & 15.7 & 39.77 & 1.91 & 0.42 & 2.16 & 2.80 & 2.24 \\
23.8 & 36.4 & -5.22 & 16.5 & 39.43 & 1.99 & 0.42 & 2.08 & 2.55 & 2.18 \\
24.5 & 34.3 & -5.23 & 17.1 & 37.88 & 2.05 & 0.42 & 2.06 & 2.49 & 2.18 \\
24.9 & 33.2 & -5.17 & 18.6 & 34.40 & 2.21 & 0.42 & 1.78 & 2.05 & 1.86 \\
25.4 & 31.9 & -5.18 & 18.9 & 34.51 & 2.23 & 0.42 & 1.82 & 2.08 & 1.91 \\
25.9 & 30.7 & -5.25 & 19.1 & 36.52 & 2.25 & 0.42 & 1.83 & 2.13 & 1.98 \\
26.4 & 29.6 & -5.35 & 19.6 & 40.63 & 2.29 & 0.42 & 1.82 & 2.15 & 2.00 \\
26.9 & 28.5 & -6.56 & 21.5 & 446.88 & 2.45 & 0.43 & 1.57 & 1.91 & 1.88 \\
27.5 & 27.2 & -6.69 & 22.0 & 684.35 & 2.50 & 0.43 & 1.56 & 1.89 & 1.86 \\
28.5 & 25.4 & -6.72 & 22.9 & 742.22 & 2.56 & 0.43 & 1.55 & 1.86 & 1.83 \\
29.1 & 24.3 & -6.48 & 23.2 & 361.98 & 2.58 & 0.43 & 1.58 & 1.99 & 1.95 \\
29.6 & 23.5 & -6.48 & 23.6 & 352.50 & 2.62 & 0.43 & 1.57 & 1.99 & 1.95 \\
30.1 & 22.7 & -6.48 & 23.9 & 355.45 & 2.64 & 0.43 & 1.59 & 1.99 & 1.95 \\
30.9 & 21.6 & -6.07 & 25.5 & 134.72 & 2.75 & 0.43 & 1.47 & 1.72 & 1.68 \\
31.6 & 20.6 & -6.12 & 26.6 & 160.77 & 2.82 & 0.43 & 1.42 & 1.69 & 1.65 \\
32.2 & 19.9 & -6.13 & 26.8 & 161.80 & 2.84 & 0.43 & 1.44 & 1.72 & 1.68 \\
33.0 & 18.9 & -5.90 & 28.0 & 94.85 & 2.92 & 0.43 & 1.38 & 1.60 & 1.56 \\
33.8 & 18.0 & -5.91 & 28.8 & 95.78 & 2.96 & 0.43 & 1.38 & 1.63 & 1.59 \\
34.5 & 17.3 & -5.91 & 29.7 & 94.79 & 3.02 & 0.43 & 1.35 & 1.62 & 1.58 \\
35.2 & 16.6 & -5.83 & 30.9 & 72.76 & 3.08 & 0.44 & 1.30 & 1.54 & 1.49 \\
36.0 & 15.9 & -6.04 & 32.1 & 101.53 & 3.15 & 0.44 & 1.26 & 1.48 & 1.44 \\
37.0 & 15.0 & -6.03 & 32.9 & 90.60 & 3.20 & 0.44 & 1.26 & 1.52 & 1.47 \\
38.0 & 14.3 & -6.12 & 33.9 & 105.48 & 3.25 & 0.44 & 1.26 & 1.52 & 1.47 \\
\hline
\end{tabular}

\end{table*}

\begin{table*}[]
    \centering
    \caption{Atmosphere fundamental properties for each model in the metallicity sequence for a fixed $T_\star = 16.1$\,kK.}
\begin{tabular}{lccccccccr}
\hline\hline
$Z/Z_\odot$ & $\log \dot{M}$ & $T_{2/3}$ & $R_{\mathrm{t}}$ & $\log g$ & $\Gamma_\mathrm{e} (R_\mathrm{crit})$ & $R_{2/3}$ & $R_{\mathrm{crit}}$ & $R_{\mathrm{sonic}}$ \\
-- & $[\mathrm{M}_\odot\,\mathrm{yr}^{-1}]$ & [kK] & [$R_{\odot}$] & [cm\,s$^{-2}$] & [$R_{\star}$] & [$R_{\star}$] & [$R_{\star}$] & [$R_{\star}$] \\
\hline
1.0 & -5.30 & 9.1 & 57.3 & 0.96 & 0.41 & 3.15 & 8.55 & 4.88 \\
0.8 & -5.32 & 9.0 & 58.1 & 0.94 & 0.41 & 3.20 & 9.21 & 5.12 \\
0.7 & -6.54 & 10.0 & 435.4 & 1.13 & 0.42 & 2.59 & 11.42 & 7.64 \\
0.6 & -6.55 & 10.0 & 441.3 & 1.12 & 0.41 & 2.61 & 11.44 & 7.64 \\
0.5 & -6.64 & 10.1 & 522.3 & 1.14 & 0.42 & 2.54 & 10.57 & 7.28 \\
0.4 & -6.91 & 10.8 & 1067.6 & 1.26 & 0.42 & 2.21 & 6.10 & 4.98 \\
0.3 & -6.93 & 11.3 & 1261.3 & 1.34 & 0.42 & 2.02 & 4.37 & 3.83 \\
0.2 & -7.06 & 11.3 & 1562.6 & 1.34 & 0.42 & 2.02 & 4.62 & 4.01 \\

\hline
\end{tabular}

    \label{tab:wind-fund-props-Z}
\end{table*}

\section{Luminosities of LBVs and cLBVs with updated distances}
\label{sec:distance-lum}

To construct the HR diagram in Fig.~\ref{fig:HRD}, we first collected literature values of effective temperatures and luminosities of OBA and yellow hypergiants, LBVs, B[e]SG, and yellow supergiants (YSGs). The list of OBA hypergiants and LBVs is largely based on previous literature compilations from \cite{Naze+2012} and \cite{Smith+2019}. The B[e]SG stars are sourced from \cite{Miroshnichenko+2025} and the YHGs and YSGs values are taken from \cite{Kasikov+2026}.
We updated the luminosities based on updated distance measurements -- mostly from the catalog of \cite{BailerJones+2021} based on Gaia DR3 parallaxes. We considered the geometric distances listed in their catalog.
In Table~\ref{tab:LBV-props-dist} we disclose the literature stellar temperatures, luminosities (with the respective distances), and updated luminosities adopting the corresponding new distances.

In general, we do not find a systematic increase in the obtained luminosities compared to previous measurements, a result that differs from what \cite{Smith+2019} found when updating previous literature values to Gaia DR2-based distances. Namely, they found an overall reduction in luminosity.
However, we find that the population of LBVs with updated distances tends to agree much more with the LBV instability strip \citep{Groh+2009}. Moreover it seems to connect smoothly with the brightest hot YSGs \citep[the edge of the so-called Yellow Void][]{deJager+1998}, whose luminosities were recently updated by \cite{Kasikov+2026}.

\onecolumn
\centering
\begin{table*}[]
\small
\caption{Literature properties and updated distances and luminosities of LBVs and OBA hypergiants.} 
\begin{tabular}{lc|p{12mm}|p{12mm}p{12mm}|p{12mm}p{12mm}|p{12mm}p{30mm}}
\hline\hline
Star {\tiny (SIMBAD)}     & Type & $T_\mathrm{eff}$\,[kK] \newline (lit.) & $\log L/\mathrm{L_\odot}$ (lit.) & $d_\mathrm{pc}$ \newline (lit.) & $\log L/\mathrm{L_\odot}$ (new) & $d_\mathrm{pc}$ \newline (new) & Ref.\,$d_\mathrm{pc}$ (new)      & Ref. \,\,\,\,\,\,\,\,\,\,\,\,\,\,\,\,\,\,\,\,\,\,\,\,\,\,\,\,\,\,\,\,\,\,\,\,     (lit.)                                        \\
\hline 
HR Car     & LBV        & 17.9 & 5.69 & 5000  & 5.63  & 4677 & BJ21   & Gr09                                      \\
           &            & 21.4 & 5.90  & 5000  & 5.84  & 4677 & BJ21   & vG01                               \\
MWC 930    & LBV        & 22.0   & 5.50  & 3500  & 6.18  & 7640 & BJ21   & Mi14                            \\
           &            & 8.0    & 5.50  & 3500  & 6.18  & 7640 & BJ21   & Mi14                            \\
HD160529   & LBV        & 10.0   & 5.46 & 2500  & 5.16  & 1769 & BJ21   & vG01                               \\
           &            & 7.8  & 5.46 & 2500  & 5.16  & 1769 & BJ21   & St91                                   \\
HD316285   & LBV        & 10.1 & 5.44 & 1850  & 5.84  & 2943 & BJ21   & Hi98                                   \\
AFGL2298   & LBV        & 11.7 & 6.17 & 10000 & 6.17  & 10000& Ue01          & Cl09a                                    \\
           &            & 10.9 & 6.11 & 10000 & 6.11  & 10000& Ue01   & Cl09a                                    \\
           &            & 10.3 & 6.30  & 10000 & 6.30  & 10000& Ue01   & Cl09a                                    \\
           &            & 10.9 & 6.17 & 10000 & 6.17  & 10000& Ue01   & Cl09a                                    \\
{[}OMN2000{]} LS1      & LBV        & 13.2 & 5.75 & 6000  & 5.44  & 4203 & BJ21   & Cl09b                                    \\
           &            & 13.4 & 5.30  & 3400  & 5.48  & 4203 & BJ21   & Cl09b                                    \\
           &            & 13.7 & 4.86 & 2000  & 5.51  & 4203 & BJ21   & Cl09b                                    \\
{\tiny Cl* Westerlund 1} W243 & LBV        & 19.0   & 5.86 & 4500  & 5.81  & 4230 & Ne22   & C-N04          \\
           &            & 8.5  & 5.86 & 4500  & 5.81  & 4230 & Ne22   & Ri09                                   \\
P Cyg      & LBV        & 18.7 & 5.78 & 1610  & 5.76  & 1574 & BJ21   & Rv20                                     \\
HD183143   & BHG        & 12.8 & 5.46 & 2200  & 5.45  & 2187 & BJ21   & W\ss22                                 \\
HD168625   & BHG        & 14.0   & 5.58 & 2800  & 5.01  & 1452 & BJ21   & Ma16                                      \\
HD168607   & BHG        & 8.5  & 5.31 & 2200  & 5.09  & 1706 & BJ21   & Ch89                                  \\
zet1 Sco   & BHG        & 17.2 & 5.93 & 1640  & 6.03  & 1845 & BJ21   & Cl12, MBP25                           \\
HD199478   & BHG        & 12.7 & 5.46 & 2400  & 5.48  & 2444 & BJ21   & W\ss23                                 \\
Schulte 12 & BHG        & 13.7 & 6.28 & 1750  & 6.21  & 1610 & BJ21   & Cl12                                     \\
BP Cru     & BHG        & 18.1 & 5.67 & 3040  & 5.81  & 3575 & BJ21   & Cl12                                     \\
HD190603   & BHG        & 18.0   & 5.58 & 1570  & 5.75  & 1918 & BJ21   & Cl12                                     \\
HD80077    & BHG        & 17.7 & 6.53 & 3600  & 6.26  & 2626 & BJ21   & Ha18                                    \\
MWC 349A   & LBV        & 25.0   & 5.75 & 1400  & 5.85  & 1563 & BJ21   & G-M12                          \\
MWC 314    & LBV        & 18.0   & 5.85 & 3000  & 6.06  & 3821 & BJ21   & Lo13                                     \\
GAL 024.73+00.69       & LBV        & 12.0   & 5.60  & 5200  & 5.26  & 3500 & Pe12     & Cl03                                     \\
Sher 25    & BSG        & 20.9 & 5.48 & 5740  & 5.50  & 5857 & BJ21   & W\ss23                                 \\
AG Car     & LBV        & 22.8 & 6.17 & 6000  & 6.02  & 5039 & BJ21   & Gr09, Gr11               \\
           &            & 22.8 & 6.17 & 6000  & 6.02  & 5039 & BJ21   & Gr09, Gr11                      \\
           &            & 21.5 & 6.17 & 6000  & 6.02  & 5039 & BJ21   & Gr09, Gr11                      \\
           &            & 17.0   & 6.17 & 6000  & 6.02  & 5039 & BJ21   & Gr09, Gr11                      \\
           &            & 16.4 & 6.00    & 6000  & 5.85  & 5039 & BJ21   & Gr09, Gr11                      \\
           &            & 14.0   & 6.00    & 6000  & 5.85  & 5039 & BJ21   & Gr09, Gr11                      \\
           &            & 14.3 & 6.04 & 6000  & 5.89  & 5039 & BJ21   & Gr09, Gr11                      \\
{[}GKF2010{]}MN 112    & LBV        & 17.4 & 6.14 & 13560 & 6.14  & 13560            & BJ21   & Mv22                                   \\
HD169454   & BHG        & 20.4 & 5.92 & 1540  & 6.12  & 1944 & BJ21   & RD22, Na95                    \\
qF 362     & LBV        & 11.3 & 6.24 & 8000  & 6.26  & 8230 & Le22         & Na09                                   \\
Pistol Star            & LBV        & 11.8 & 6.20  & 8000  & 6.22  & 8230 & Le22         & Na09                                   \\
GRS 79.29+0.46         & LBV        & 20.4 & 5.40  & 1700  & 5.54  & 2000 & Hi94 & Ag14                                 \\
WRAY 17-96 & LBV        & 20.0   & 6.47 & 4500  & 6.45  & 4383 & BJ21   & Eg02                                      \\
Peony Star & WNL        &       25.0   & 6.50  & 8000  & 6.52  & 8230 & Le22         & Ba08                                  \\
eta Car    & LBV        & 9.4  & 6.69 & 2000  & 6.85  & 2400 & Hu23           & Gr12                                      \\
Hen 3-519  & LBV        & 15.1 & 6.26 & 8000  & 6.19  & 7342 & BJ21   & S-S17, vG01 \\
{[}GKF2010{]} MN48     & LBV        & 14.0   & 5.50  & 4000  & 5.58  & 4369 & BJ21   & Kn16                                   \\
AS 314     & PAGB       & 9.5  & 3.65 & 1614  & 3.65  & 1614 & BJ21   & Bh25                                \\
WRAY 16-232            & LBV        & 12.5 & 5.70  & 2100  & 5.75  & 2226 & BJ21   & Ar25                                      \\
GAL 026.47+00.02       & LBV        & 17.0   & 6.00    & 6500  & 5.97  & 6267 & BJ21   & Cl03                                     \\
GCIRS 16NE & WNL        & 24.1 & 6.34 & 8500  & 6.31  & 8230 & Le22         & Nz12, Na97           \\
GCIRS 16C  & WNL        & 19.5 & 5.90  & 7600  & 5.97  & 8230 & Le22         & Ma07                                   \\
GCIRS 33SE & WNL        & 18.0   & 5.75 & 7600  & 5.82  & 8230 & Le22         & Nz12                                      \\
GCIRS 16SW & WNL        & 24.4 & 6.41 & 8500  & 6.38  & 8230 & Le22         & Na97                                   \\
WRAY 15-751            & LBV        & 30.0   & 5.91 & 6000  & 6.10  & 7444 & BJ21   & Nz12, Pa06, vG01     \\
GCIRS 34W  & WNL        & 19.5 & 5.50  & 7600  & 5.57  & 8230 & Le22         & Ma07                                   \\
GS Mus     & OHG        & 28.0   & 5.28 & 2784  & 5.28  & 2784 & BJ21   & This work                                   \\
HD151804   & OHG        & 28.2 & 6.18 & 2140  & 6.18  & 2140 & BJ21   & This work             \\      
\hline
\end{tabular}
\vspace{-0.2cm}
\tablefoot{Gr09: \cite{Groh+2009}, vG01: \cite{vanGenderen2001}, Mi14: \cite{Miroshnichenko+2014}, St91: \cite{Sterken+1991}, Hi98: \cite{Hillier+1998}, Cl09a: \cite{Clark+2009a}, Cl09b: \cite{Clark+2009b}, C-N04: \cite{Clark-Negueruela2004}, Ne22: \cite{Negueruela+2022}, Ri09: \cite{Ritchie+2009}, Rv20: \cite{Rivet+2020}, W\ss22: \cite{Wessmayer+2022}, Ma16: \cite{Mahy+2016}, Ch89: \cite{Chentsov+1989}, Cl12: \cite{Clark+2012}, BP25: \cite{Bernini-Peron+2025}, W\ss23: \cite{Wessmayer+2023}, Ha18: \cite{Haucke+2018}, G-M12: \cite{Gvaramadze-Menten2012}, Lo13: \cite{Lobel+2013}, Cl03: \cite{Clark+2003}, W\ss23: \cite{Wessmayer+2023}, Gr11: \cite{Groh+2011}, Mv22: \cite{Maryeva+2022}, RD22: \cite{RubioDiez+2022}, Na95: \cite{Najarro1995}, Na09: \cite{Najarro+2009}, Ag14: \cite{Agliozzo+2014}, Eg02: \cite{Egan+2002}, Ba08: \cite{Barniske+2008}, Gr12: \cite{Groh+2012}, S-S17: \cite{Smith-Stassum2017}, Kn16: \cite{Kniazev+2016}, Bh25: \cite{Bakhytkyzy+2025}, Ar25: \cite{Arun+2025}, Nz12: \cite{Naze+2012}, Ma07: \cite{Martins+2007}, Na97: \cite{Najarro+1997}, Ue01: \cite{Ueta+2001}. Pa06: \cite{Pasquali+2006}, Sm06: \cite{Smith2006}, Hu23: \cite{Hur+2023}, Le22:\cite{Leung+2023}, Pe12: \cite{Petriella+2012}}
\label{tab:LBV-props-dist}

\end{table*}

\end{appendix}

\end{document}